\documentclass{article}
\usepackage{amsbsy}
\usepackage{amssymb}
\usepackage{amstext}

\newcommand{\be}{\begin{equation}}
\newcommand{\ee}{\end{equation}}
\newcommand{\bea}{\begin{eqnarray}}
\newcommand{\eea}{\end{eqnarray}}
\newcommand{\ben}{\begin{eqnarray*}}
\newcommand{\een}{\end{eqnarray*}}
\newcommand{\dst}{\displaystyle}

\newcommand{\pa}{\partial}

\newcommand{\ve}{\varepsilon}

\def\fko#1{\stackrel{\scriptscriptstyle(#1)}{f_o}}
\def\fkg#1{\stackrel{\scriptscriptstyle(#1)}{f_g}}

\begin{document}

\title{Data analysis of gravitational-wave signals
\\from spinning neutron stars.
\\II.\ Accuracy of estimation of parameters}

\author{Piotr Jaranowski$^{1,2}$
\and Andrzej Kr\'olak$^{1,3}$
\\{\it $^1$Albert-Einstein-Institut,
       Max-Planck-Institut f\"ur Gravitationsphysik}\\
  {\it Schlaatzweg 1, 14473 Potsdam, Germany}
\\{\it $^2$Institute of Physics, Bia{\l}ystok University}\\
  {\it Lipowa 41, 15-424 Bia{\l}ystok, Poland}
\\{\it $^3$Institute of Mathematics,
       Polish Academy of Sciences}\\
  {\it \'{S}niadeckich 8, 00-950 Warsaw, Poland}}
\maketitle

\begin{abstract}

We examine the accuracy of estimation of parameters of the
gravitational-wave signals from spinning neutron stars that can be achieved from
observations by Earth-based laser interferometers.  We consider a model of the
signal consisting of two narrowband components and including both phase and
amplitude modulation.  We calculate approximate values of the rms errors of the
parameter estimators using the Fisher information matrix.  We carry out
extensive Monte Carlo simulations and obtain cumulative distribution functions
of rms errors of astrophysically interesting parameters:  amplitude of the
signal, wobble angle, position of the source in the sky, frequency, and spindown
coefficients.  We consider both all-sky searches and directed searches.  We also
examine the possibility of determination of neutron star proper motion.  We
perform simulations for all laser-interferometric detectors that are currently
under construction and for several possible lengths of the observation time and
sizes of the parameter space.  We find that observations of continuous
gravitational-wave signals from neutron stars by laser-interferometric detectors
will provide a very accurate information about their astrophysical properties.
We derive several simplified models of the signal that can be used in the
theoretical investigations of the data analysis schemes independently of the
physical mechanisms generating the gravitational-wave signal.

\end{abstract}

\section{Introduction}

Detection of gravitational waves from spinning neutron stars by currently
constructed long-arm laser interferometers \cite{GEO600,LIGO,VIRGO,TAMA300} is
expected to provide a wealth of astrophysical information about these objects.
In the first paper of this series \cite{P1} (hereafter Paper I) we introduced a
model of the gravitational-wave signal from a spinning neutron star.  The
response of a laser interferometer to such a signal will carry information about
the neutron star period of rotation and its evolution, amplitude of the signal
and its polarization.  Moreover even the response of a single detector will
contain the full information about the position of the source in the sky.  Also
higher order terms in the phase contain velocity of the source and its distance.
In this second paper of our series we investigate in detail how accurately these
astrophysically interesting parameters can be determined from observations by
laser-interferometric detectors.  We assume that optimal data analysis methods
presented in Paper I are used.  We use a standard tool, the Fisher information
matrix, to assess the rms errors of our estimation method.

We find impressive potential accuracies achievable with gravitational-wave
detectors that can make them into astronomical laboratories competitive with
optical and radio observatories.  For all-sky searches and 120 days of
observation time the initial laser-interferometric detectors will be able to
locate the strongest fast and young neutron star gravitational-wave emitters at
a distance of 1 kpc with accuracy of order $10^{-7}$ sr and estimate the
frequency and the four spindown parameters of the wave with relative accuracies
of order $10^{-2(4-k)}$, where $k=0,\ldots,4$ denotes the $k$th spindown ($k=0$
corresponds to the frequency of the wave).  Amplitude can be estimated within a
few percent and wobble angle within a few percent of a radian.  For directed
searches the relative rms errors of estimation of frequency and spindowns will
decrease with respect to the all-sky searches by a factor of $10$ for the fourth
spindown, $10^2$ for the third and the second spindowns, $10^3$ for the first
spindown, and $10^4$ for the frequency.  With advanced detectors a reasonable
accuracy of estimation of proper motions of neutron stars with all-sky searches
will be achievable for nearby neutron stars (at distances of 40 pc or so).

The plan of the paper is as follows.  In Sec.\ 2 we review the model of the
gravitational-wave signal introduced in Paper I.  We introduce eight special
cases of the model depending on whether observation time is long or short and
neutron star is young or old and it is spinning fast or slow.  In Sec.\ 3 we
present calculations of the rms errors that can be achieved for our models
assuming that we use maximum likelihood detection as data analysis method.  To
estimate the rms errors we use Fisher information matrix.  In Sec.\ 4 we
explore the possibility of estimating proper motion of neutron stars from
gravitational-wave observations.  In Sec.\ 5 we introduce a number of
simplified signal models that enable us to assess the rms errors of the phase
parameters independently of the physical mechanisms generating the gravitational
radiation from a spinning neutron star.  In Sec.\ 6 using a simplified model
we investigate the dependence of the rms errors of the parameters on the
observational parameters:  duration of observation, initial moment of
observation and the latitutde of the detector.

Several additional topics are investigated in the appendices.  In Appendix A we
interpret our 1/4 of a cycle criterion in terms of the signal-to-noise ratio
loss.  In Appendix B we give details of number of cycles calculations for our
special models.  In Appendix C we study the polynomial phase model and the
effect of the choice of the initial time in the parameter estimation problem.

\section{A model of the two-component gravitational-wave signal}

In Sec.\ II of Paper I we have introduced the following two-component model of 
the gravitational-wave signal $h$ from a spinning neutron star:
\be
\label{s1}
h(t) = h_1(t) + h_2(t),
\ee
with
\bea
\label{s2}
&h_1(t) = F_+(t) h_{1+}(t) + F_\times(t) h_{1\times}(t),\quad
h_2(t) = F_+(t) h_{2+}(t) + F_\times(t) h_{2\times}(t),&\\
\label{s3}
&h_{1+}(t) = \frac{1}{8}h_o\sin2\theta\sin2\iota\cos\Psi(t),\quad
h_{2+}(t) = \frac{1}{2}h_o\sin^2\theta(1+\cos^2\iota)\cos2\Psi(t),&\\
\label{s4}
&h_{1\times}(t) = \frac{1}{4}h_o\sin2\theta\sin \iota\sin\Psi(t),\quad
h_{2\times}(t) = h_o\sin^2\theta\cos \iota\sin2\Psi(t).&
\eea
The model of the signal given above represents the quadrupole gravitational wave
that is emitted by a freely precessing axisymmetric star.  The angle $\theta$,
called the wobble angle, is the angle between the total angular momentum vector
of the star and the star's axis of symmetry and $\iota$ is the angle between the
total angular momentum vector of the star and the direction from the star to the
Earth.  For the case when $\theta=\pi/2$ (then the component $h_1$ vanishes) the
model also gives the quadrupole wave from a triaxial ellipsoid rotating about a
principal axis.  In both cases the amplitude $h_o$ is given by
\be
\label{ho}
h_o = \frac{16\pi^2G}{c^4}\frac{\epsilon I f_o^2}{r_o},
\ee
where $I$ is the moment of inertia with respect to the rotation axis and 
$r_o$ is the distance to the star. For precessing axisymmetric star 
$f_o$ is the sum of the frequency of rotation of the
star and the frequency of precession and $\epsilon$ is the poloidal 
ellipticity of the star whereas for triaxial ellipsoid 
$f_o$ is the frequency of rotation of the star and $\epsilon$ is ellipticity
of the star defined by $\epsilon:=(I_1 - I_2)/I$,
where  $I_1$ and $I_2$ are the moments of inertia of the star with respect to 
the principal axes orthogonal to the rotation axis. 

Inserting numerical values into Eq.\ (\ref{ho}) gives
\be
\label{hon}
h_o = 4.23\times10^{-25}d_o\left(\frac{f_o}{\rm 100~Hz}\right)^2,
\ee 
where
\be
\label{do}
d_o := \frac{\epsilon}{10^{-5}}\,\frac{I}{10^{45}~\mbox{g cm}^2}
\,\frac{\mbox{1~kpc}}{r_o}.
\ee 
In all numerical simulations presented in the next sections except for
the case when the effect of the proper motion was studied we had assumed 
the numerical value of the parameter $d_o$ to be $1$. The most uncertain 
parameter in Eq.\ (\ref{do}) is the ellipticity $\epsilon$, the value of 
$\epsilon\sim10^{-5}$ is thought to be an upper limit and typical values may be 
significantly less \cite{BCCS97}.

The beam-pattern functions $F_+$ and $F_\times$ from Eq.\ (\ref{s2}) are given 
by 
\bea
\label{bp5a}
F_+(t) &=& \sin\zeta\left[a(t)\cos2\psi+b(t)\sin2\psi\right],\\
\label{bp5b}
F_\times(t) &=& \sin\zeta\left[b(t)\cos2\psi-a(t)\sin2\psi\right],
\eea
where
\bea
\label{adef}
a(t) &=&
\frac{1}{16}\sin2\gamma(3 - \cos2\lambda)(3 - \cos2\delta)
\cos[2(\alpha-\phi_r-\Omega_r t)]
\nonumber\\&&
-\frac{1}{4}\cos2\gamma\sin\lambda(3 - \cos2\delta)
\sin[2(\alpha-\phi_r-\Omega_r t)]
\nonumber\\&&
+\frac{1}{4}\sin2\gamma\sin2\lambda\sin2\delta
\cos[\alpha-\phi_r-\Omega_r t]
\nonumber\\&&
-\frac{1}{2}\cos2\gamma\cos\lambda\sin2\delta
\sin[\alpha-\phi_r-\Omega_r t]
\nonumber \\&&
+\frac{3}{4}\sin2\gamma\cos^2\lambda\cos^2\delta,
\\
\label{bdef}
b(t) &=&
\cos2\gamma\sin\lambda\sin\delta\cos[2(\alpha-\phi_r-\Omega_r t)]
\nonumber \\&&
+\frac{1}{4}\sin2\gamma(3 - \cos2\lambda)\sin\delta
\sin[2(\alpha-\phi_r-\Omega_r t)]
\nonumber \\&&
+\cos2\gamma\cos\lambda\cos\delta\cos[\alpha-\phi_r-\Omega_r t]
\nonumber \\&&
+\frac{1}{2}\sin2\gamma\sin2\lambda\cos\delta
\sin[\alpha-\phi_r-\Omega_r t].
\eea
The angles $\alpha$, $\delta$, and $\psi$ are respectively right ascension,
declination of the gravitational-wave source, and polarization angle (these
angles determine the orientation of the wave reference frame with respect to the
celestial sphere reference frame).  The angle $\lambda$ is the latitude of the
detector's site, $\gamma$ determines the orientation of the detector's arms with
respect to local geographical directions and $\zeta$ is the angle between the
interferometer arms.  $\Omega_r$ is the rotational angular velocity of the
Earth, and $\phi_r$ is a deterministic phase which defines the position of the
Earth in its diurnal motion at $t=0$. The derivation of the formulae 
(\ref{bp5a})--(\ref{bdef}) is given in Sec.\ II~A of Paper I.

The phase $\Psi$ of Eqs.\ (\ref{s3}) and (\ref{s4}) is given by
\be
\label{ap200}
\Psi(t) = \Phi_0 + 2\pi \sum_{k=0}^{s_1}{\fko k}\frac{t^{k+1}}{(k+1)!}
+ \frac{2\pi}{c} {\bf n}_0\cdot{\bf r}_{\rm ES}(t)
\sum_{k=0}^{s_2}{\fko k}\frac{t^k}{k!}
+ \frac{2\pi}{c} {\bf n}_0\cdot{\bf r}_{\rm E}(t)
\sum_{k=0}^{s_3}{\fko k}\frac{t^k}{k!},
\ee
where ${\bf r}_{\rm ES}$ is the vector joining the solar system barycenter (SSB)
with the center of the Earth and ${\bf r}_{\rm E}$ joins the center of the Earth
with the detector, ${\fko k}$ is the $k$th time derivative of the instantaneous
frequency at the SSB evaluated at $t = 0$, ${\bf n}_0$ is the constant unit
vector in the direction of the star in the SSB reference frame.  We have
neglected the relativistic effects.  To derive the above model we assumed that
in the rest frame of the neutron star the gravitational-wave frequency can be
expanded in a power series.  For the detailed derivation of the phase model 
(\ref{ap200}) see Sec.\ II~B and Appendix A of Paper I.

Neglecting the eccentricity of the Earth's orbit and the motion of the Earth
around the Earth-Moon barycenter the scalar products ${\bf n}_0\cdot{\bf r}_{\rm 
E}$ and ${\bf n}_0\cdot{\bf r}_{\rm ES}$ of Eq.\ (\ref{ap200}) can be written as 
(see Sec.\ II~B of Paper I) 
\bea
\label{n0re}
{\bf n}_0\cdot{\bf r}_{\rm E} &=& R_{E}\left[\sin\lambda\sin\delta
+\cos\lambda\cos\delta\cos\left(\alpha-\phi_r-\Omega_r t\right)\right],
\\
\label{n0res}
{\bf n}_0\cdot{\bf r}_{\rm ES} &=& 
R_{ES}\left[\cos\alpha\cos\delta\cos\left(\phi_o+\Omega_o t\right)
+\left(\cos\ve\sin\alpha\cos\delta+\sin\ve\sin\delta\right)
\sin\left(\phi_o+\Omega_o t\right)\right],
\eea
where $R_{ES} =$ 1 AU is the mean distance from the Earth's center to the SSB,
$R_E$ is the mean radius of the Earth, $\Omega_o$ is the mean orbital angular
velocity of the Earth, and $\phi_o$ is a deterministic phase which defines the
position of the Earth in its orbital motion at $t=0$, $\ve$ is the angle between 
ecliptic and the Earth's equator.

The extremal values of the spindown parameters ${\fko k}$ are estimated from the
formula
\be
\label{extr}
{\fko k} \simeq (-1)^k k!\frac{f_o}{\tau^k},\quad k=1,2,\ldots,
\ee
where $\tau$ we call the spindown age of the star.

We choose the number of terms in the power series on the right-hand side of Eq.\
(\ref{ap200}) according to the following criterion:  {\em we exclude an effect
from the model of the signal in the case when it contributes less than 1/4 of a
cycle to the phase of the signal during the observation time}.  An
interpretation of the 1/4 of a cycle criterion in terms of the signal-to-noise
ratio loss is given in Appendix A.  If we restrict to observation times
$T_o\le120$ days, frequencies $f_o\le1000$ Hz, and spindown ages $\tau\ge40$
years, the phase model (\ref{ap200}) meets the criterion for an appropriate
choice of the numbers $s_1$, $s_2$, and $s_3$.  Also the effect of the star
proper motion in the phase is negligible if we assume that the
star moves w.r.t.\ the SSB not faster than $10^3$ km/s and its distance to the
Earth $r_o\ge1$ kpc.  However for nearby neutron stars at distances $r_o\ge40$
pc the proper motion contribution to the phase (\ref{ap200}) can exceed 1/4 of a
cycle.  We discuss this issue at the end of the present section.

We consider several models depending on the range of the gravitational-wave
frequency $f_g$ [$f_g=f_o$ for the $h_1$ component and $f_g=2f_o$ for the $h_2$
component of the signal (\ref{s1})], spindown age $\tau$, and observation time
$T_o$.  Following Brady {\it et al}.\ \cite{BCCS97} we say that neutron star is
{\em slow} if $f_g\simeq 200$ Hz and {\em fast} if $f_g\simeq 1$ kHz.  We define
a neutron star to be {\em old} if $\tau \simeq 10^3$ yr and {\em young} if $\tau
\simeq 40$ yr.  We consider observation time to be {\em short} if $T_o \simeq 7$
days and {\em long} if $T_o \simeq 120$ days.  Accordingly we have eight
possible models.  In Table 1 we present the number of terms needed in the power
series of various contributions to the phase in order to meet the above
criterion.  The details of the calculation are given in Appendix B.

\begin{table}
\begin{center}
\begin{tabular}{|c|c|c|c|c|c|}\hline
$T_o$ (days) & $\tau_{\rm min}$ (years) & $f_{g{\rm max}}$ (Hz) & 
$s_1$  & $s_2$  & $s_3$ \\ \hline\hline
120 & 40     & $10^3$ & 4 & 3 & 0 \\ \hline
120 & 40     & 200    & 4 & 2 & 0 \\ \hline
120 & $10^3$ & $10^3$ & 2 & 1 & 0 \\ \hline
120 & $10^3$ & 200    & 2 & 1 & 0 \\ \hline\hline
7   & 40     & $10^3$ & 2 & 1 & 0 \\ \hline
7   & 40     & 200    & 2 & 1 & 0 \\ \hline
7   & $10^3$ & $10^3$ & 1 & 1 & 0 \\ \hline
7   & $10^3$ & 200    & 1 & 1 & 0 \\ \hline
\end{tabular}
\end{center}
\caption{The number of spindown terms needed in various contributions to the
phase of the signal depending on the type of population of neutron stars
searched for.  The number $s_1$ refers to the spindown contribution, $s_2$ 
refers to the Earth orbital motion, and $s_3$ refers to the Earth diurnal motion 
contribution [cf.\ Eq.\ (\ref{ap200})].}
\end{table}

In our calculations of the rms errors of the estimators of the parameters we
have also studied the case of directed searches.  By directed searches we mean
the search for the signal from a spinning neutron star which position in the sky
is known.  This reduces the dimension of the parameter space by 2 since the
right ascension and the declination of the source are known.  We do not assume
that the gravitational wave frequency is known. Even if we are searching for a
gravitational wave from a known pulsar its frequency may not coincide with the
frequency of the electromagnetic radiation or be exactly twice that frequency.
For directed searches the values of the parameters $s_1$, $s_2$, and $s_3$ given
in Table 1 remain in general the same however for specific positions in the sky
they may be less.

The numbers of spindowns that must be included in the phase model given in Table
1 agree with the corresponding numbers of Brady {\it et al}.\ \cite{BCCS97}
obtained by a different criterion for the cases they considered (see Figure 6 of
their paper).

We have also considered the effect of the proper motion of the neutron star on
the phase of the signal assuming that it moves uniformly with respect to the SSB
reference frame.  We have found that for the observation time $T_o=$ 120 days
and the extreme case of a neutron star at a distance $r_o=$ 40 pc, transverse
velocity $|{\bf v}_{{\rm ns}\perp}|=10^3$ km/s (where ${\bf v}_{{\rm ns}\perp}$
is the component of the star's velocity ${\bf v}_{\rm ns}$ perpendicular to the
vector ${\bf n}_{0}$), gravitational-wave frequency $f_g=2f_o=$ 1 kHz, and
spindown age $\tau=$ 40 years, proper motion contributes only $\sim$4 cycles to
the phase of the signal.  This dominant proper motion contribution $\Psi_{\rm
pm}$ to the phase (\ref{ap200}) is given by (cf.\ Eq.\ (A20) in Appendix A of 
Paper I)
\be
\label{pmphase}
\Psi_{\rm pm}(t) = \frac{2\pi}{c}
\frac{{\bf v}_{{\rm ns}\perp}\cdot{\bf r}_{\rm ES}(t)}{r_o}f_o\,t.
\ee
The ratio ${\bf v}_{{\rm ns}\perp}/r_o$ determines the proper motion of the 
star.  We shall study the accuracy of estimation of the proper motion in Sec.\ 
4.

\section{Estimation of parameters of the two-component signal}

In Sec.\ III of Paper I we have presented an optimum data analysis method:  {\em
maximum likelihood detection}.  This method consists of maximizing the
likelihood function $\Lambda$ with respect to the parameters of the signal.  If
the maximum of $\Lambda$ exceeds a certain threshold calculated from the false
alarm probability that we can afford we say that the signal is detected.  The
values of the parameters that maximize $\Lambda$ are said to be the {\em maximum
likelihood estimators} of the parameters of the signal.  Let us collect the
signal parameters into the vector
$\boldsymbol{\theta}=(\theta_1,\ldots,\theta_n)$.  The covariance matrix $C$ of
the estimators of the parameters $\boldsymbol{\theta}$ is approximately given by
the inverse of the {\em Fisher information matrix $\Gamma$}:
$C\cong\Gamma^{-1}$.  This approximation is the better the higher the
signal-to-noise ratio.  For a signal $h$ buried in a Gaussian noise the
components of the Fisher matrix $\Gamma$ are given by
\be
\Gamma_{ij} =
\left(\frac{\partial{h}}{\partial{\theta_i}}\Bigg\vert
\frac{\partial{h}}{\partial{\theta_j}}\right),
\ee
where the scalar product $(\,\cdot\,|\,\cdot\,)$ is defined by 
\be
\label{SP}
(x|y):=
4\Re\int^{\infty}_{0}\frac{\tilde{x}(f)\tilde{y}^{*}(f)}{S_h(f)}df,
\ee
where $\tilde{}$ denotes the Fourier transform, $*$ is complex conjugation,
and $S_h$ is the {\em one-sided} spectral density of the detector's noise.

For a signal $h=h_1+h_2$ that consists of two narrowband components around the
frequencies $f_o$ ($h_1$ component) and $2f_o$ ($h_2$ component), assuming that
over the bandwidth of the signal the noise spectral density $S_h$ is nearly
constant and equal to $S_h(f_o)$ or $S_h(2 f_o)$, the Fisher matrix $\Gamma$ is
approximately given by
\be
\label{gammaij}
\Gamma_{ij} \cong 
\frac{2}{S_h(f_o)} \int^{To/2}_{-To/2}
\frac{\partial{h_1}}{\partial{\theta_i}}
\frac{\partial{h_1}}{\partial{\theta_j}}dt +
\frac{2}{S_h(2f_o)}\int^{To/2}_{-To/2}
\frac{\partial{h_2}}{\partial{\theta_i}}
\frac{\partial{h_2}}{\partial{\theta_j}}dt,
\ee
where $T_o$ is the observation time and the observation interval is 
$\left[-T_o/2,T_o/2\right]$.

For the two-component signal defined by Eqs.\ (\ref{s1})--(\ref{s4}) we can 
extract explicitely the dependence of the elements of the covariance matrix $C$
on the amplitude parameter $h_o$.  Let us write the signal (\ref{s1}) in the 
form
\be
\label{s5}
h(t;\boldsymbol{\theta})
= h_o\left[\chi_1(t;\boldsymbol{\zeta}) + \chi_2(t;\boldsymbol{\zeta})\right],
\ee
where $\boldsymbol{\theta}=(h_o,\boldsymbol{\zeta})$, i.e.\ $\boldsymbol{\zeta}$ 
is the vector of all the signal parameters with the exception of $h_o$. Let us 
note that $\boldsymbol{\zeta} = 
(\alpha,\delta,\psi,\iota,\theta,\Phi_0,f_o,{\fko 1},\ldots,{\fko s})$, so the 
signal (\ref{s5}) depends on $8+s$ parameters, where $s$ is the number of the 
spindowns one wants to include.
Substituting Eq.\ (\ref{s5}) into (\ref{gammaij}) one can 
show that the components of the covariance matrix $C\cong\Gamma^{-1}$ can be 
written using the matrix notation as (superscript $T$ denotes matrix 
transposition):
\bea
\label{cov1}
C_{h_o h_o} &\cong& K, \\
C_{h_o\boldsymbol{\zeta}} &\cong& -\frac{1}{h_o} K \Delta'(\Delta'')^{-1}, \\
C_{\boldsymbol{\zeta}\boldsymbol{\zeta}} &\cong& \frac{1}{h_o^2}
\left[(\Delta'')^{-1} + 
K(\Delta'')^{-1}(\Delta')^{T}\Delta'(\Delta'')^{-1}\right],
\eea
where
\bea
K &:=& \left[\Delta-\Delta'(\Delta'')^{-1}(\Delta')^{T}\right]^{-1},\\
\Delta &:=& \frac{2}{S_h(f_o)}
\int^{To/2}_{-To/2} \left[\chi_1(t;\boldsymbol{\zeta})\right]^2 dt
+ \frac{2}{S_h(2f_o)}
\int^{To/2}_{-To/2} \left[\chi_2(t;\boldsymbol{\zeta})\right]^2 dt,\\
\Delta'_i &:=& \frac{2}{S_h(f_o)}\int^{To/2}_{-To/2}\chi_1(t;\boldsymbol{\zeta})
\frac{\partial{\chi_1(t;\boldsymbol{\zeta})}}{\partial{\zeta_i}} dt +
\frac{2}{S_h(2 f_o) }\int^{To/2}_{-To/2}\chi_2(t;\boldsymbol{\zeta})
\frac{\partial{\chi_2(t;\boldsymbol{\zeta})}}{\partial{\zeta_i}} dt, \\
\label{cov7}
\Delta''_{ij} &:=& \frac{2}{S_h(f_o) }\int^{To/2}_{-To/2}
\frac{\partial{\chi_1(t;\boldsymbol{\zeta})}}{\partial{\zeta_i}}
\frac{\partial{\chi_1(t;\boldsymbol{\zeta})}}{\partial{\zeta_j}} dt +
\frac{2}{S_h(2 f_o) }\int^{To/2}_{-To/2}
\frac{\partial{\chi_2(t;\boldsymbol{\zeta})}}{\partial{\zeta_i}}
\frac{\partial{\chi_2(t;\boldsymbol{\zeta})}}{\partial{\zeta_j}} dt.
\eea

The optimal signal-to-noise ratio $d:=\sqrt{(h|h)}$ for the signal (\ref{s5}) 
can be computed from the formula (cf.\ Sec.\ III~C of Paper I):
\be
\label{snr}
d^2 \cong \frac{2h_o^2}{S_h(f_o)}
\int^{To/2}_{-To/2} \left[\chi_1(t;\boldsymbol{\zeta})\right]^2 dt
+ \frac{2h_o^2}{S_h(2f_o)}
\int^{To/2}_{-To/2} \left[\chi_2(t;\boldsymbol{\zeta})\right]^2 dt.
\ee

We have used the formulae (\ref{cov1})--(\ref{cov7}) to calculate numerically
covariance matrices for different phase models. Because the components of the 
covariance matrices
depend in a complicated way on the trigonometric functions of the angles 
$\alpha$, $\delta$, $\psi$, $\iota$, and $\theta$ we have resorted to the
Monte Carlo simulations. For each simulation of a covariance matrix we have 
generated 1000 sets of the angles $\{\alpha,\delta,\psi,\iota,\theta\}$ 
according to the probability measure
$$
d\alpha \times d\sin\delta\times d\psi \times d\cos \iota \times d\theta
$$
defined on the parameter space
$$
\{ \alpha\in[0,2\pi), \sin\delta\in[-1,1], 
\psi\in[0,2\pi), \cos\iota\in[-1,1], \theta\in[0,\pi] \}.
$$

In the simulations we have assumed the following models of the noise spectral
densities $S_h$ in the individual detectors.  The noise curves for the VIRGO and
the initial/advanced LIGO detectors are adopted from \cite{S96}, and the noise
curve for the TAMA300 detector is taken from \cite{TAMA300,SKpc}.  Wideband and
narrowband versions of the GEO600 detector noise are based on \cite{KSpc}.

We present results of the simulations by plotting the cumulative distribution
functions for the square roots of the diagonal components of the covariance
matrix which approximate the rms errors of the estimators of the parameters.
For some cases we also give tables with quartiles of the distributions
\cite{quartiles}.  We study the rms errors (denoted by $\sigma$) of the
following astrophysically important parameters:  amplitude $h_o$, wobble angle
$\theta$, position of the source on the sky, frequency $f_o$, and spindown 
parameters ${\fko
k}$.  The accuracy of the position on the sky is measured by the solid angle
$\Delta\Omega$ corresponding to the ellipse of semiaxes $\sigma(\delta)$ and
$\sigma(\alpha)$ and is given by
\be
\Delta\Omega = \pi\cos\delta\sigma(\delta)\sigma(\alpha).
\ee
To assess the relative rms errors of the spindown parameters ${\fko k}$ we have 
assumed in the simulations the extremal values of these parameters
given by formula (\ref{extr}).

In Figure 1 we have given cumulative distributions for the rms errors of the
parameters for all the laser-interferometric detectors under construction.  For
the GEO600 detector we have also considered the narrowband configuration tuned
to 1 kHz with bandwidth of 30 Hz.  We have taken long observation time
($T_o=120$ days) and we have assumed the neutron star to be fast ($f_o=500$ Hz)
and young ($\tau=40$ yr).  In addition in Table 2 we have given the quartiles of
the distributions plotted in Figure 1.  We observe that the rms errors for the
case of advanced detectors are 10 times less than for the initial detectors.
This is a result of simplified assumptions for the spectral density of shot
noise for the advanced detectors, namely it is taken exactly 100 times less than
for the initial detectors (for frequency range of 500 Hz to 1 kHz the shot noise
dominates).

The GEO600 detector in narrowband configuration tuned to a certain frequency (1
kHz in this case) is able to achieve accuracies of estimation of parameters in
the phase of the signal (position on the sky, frequency and spindown parameters)
several times better than kilometer-length initial detectors.  This is done at
the expense of loss of information at other frequency bands.  We see that the
accuracy of the estimation of the amplitude $h_o$ and the angle $\theta$ for the
narrowband case is less than for any other detector.  This is because the
narrowband GEO600 detector will see mainly one component of the signal ($h_2$ in
this case) where parameters $h_o$ and $\theta$ degenerate
into one parameter.  The information about the other component is small
resulting in a small accuracy for joint estimation of parameters $h_o$ and
$\theta$.  It implies that for the narrowband GEO600 detector it is more
suitable to model the signal as consisting only of the second ($h_2$) component.
Then the parameters $h_o$ and $\theta$ merge into an effective amplitude
parameter $h'_o=h_o\sin^2\theta$ [cf.\ Eqs.\ (\ref{s1})--(\ref{s4})].  We have
performed the Monte Carlo simulations for this case.  On the left top panel in
Figure 1 we have shown the cumulative distribution function for the relative rms
error of the amplitude $h'_o$.  We have checked that the position error
$\Delta\Omega$ and the spindown relative errors $\sigma({\fko k})/\vert{\fko
k}\vert$ ($k=0,\ldots,4$) are almost the same as for the case of the
two-component signal model indicating negligible contribution of the first
component ($h_1$) to the information about the phase parameters.

From Table 2 we see that the initial laser-interferometric gravitational-wave
detectors will be able to locate the strongest fast and young neutron star
gravitational-wave emitters at a distance of 1 kpc with accuracy of order
$10^{-7}$ sr and estimate the relative accuracies of frequency ($k = 0$) and
four spindowns ($k=1,\ldots,4$) within a factor of order $10^{-2(4-k)}$.
Amplitude can be estimated within a few percent and angle $\theta$ within a few
percent of a radian.

In Figures 2 and 3 we have performed the Monte Carlo simulations of the rms
parameter errors for the eight models of the phase summarized in Table 1 above.
Figure 2 is for long observation time $T_o=120$ days and Figure 3 is for short
observation time $T_o=7$ days.  Because the amplitude of the signal for our
model increases as square of the frequency [see Eq.\ (\ref{ho})] the rms errors
of the parameters for fast rotating stars are smaller than for slowly rotating
ones.  For the case of young neutron stars the spindowns are larger than for the
old ones [see Eq.\ (\ref{extr})] and this leads to more spindown terms in the
phase for young stars in order to meet the 1/4 of a cycle criterion (see Table
1).  Since spindown parameters are strongly correlated (see Appendix C)
increasing their number decreases the accuracy of their estimation.  Thus
accuracies for estimation of parameters for young neutron stars tend to be less
than for the old ones.  We observe that the accuracy of estimation of the phase
parameters (except for the fourth spindown ${\fko 4}$) is in all cases very
good.  The distributions of the rms errors of the amplitude parameters $h_o$ and
$\theta$ for various distinct models of the phase in the case when the frequency
$f_o$ and consequently the amplitude $h_o$ is the same are indistinguishable
(see top panels of Figures 2 and 3).  This shows that the amplitude parameters
and the phase parameters in our signal model are effectively uncorrelated.

In Figures 4--6 and Table 3 we have presented results of the same simulations as
in Figures 1--3 and Table 2 but for directed searches.  Here because there are
two unknown parameters less in the phase of the signal (the right ascension
$\alpha$ and the declination $\delta$ are assumed to be known) the accuracy of
estimation of the phase parameters increases.  For the initial detectors the
relative rms errors of the frequency ($k=0$) and the first two spindowns
($k=1,2$) are of the order of $10^{-3(4-k)}$, and the relative rms errors of the
third and the fourth spindown are of the order $10^{-4}$ and $10^{-1}$,
respectively.  This gives a decrease with respect to the all-sky searches by
factors of $10^4$, $10^3$, $10^2$, $10^2$, and $10$ for the frequency, the 
first,
the second, the third, and the fourth spindown respectively.  
Again the distributions of the
rms errors of the amplitude parameters $h_o$ and $\theta$ for the models with
the same frequency $f_o$ are indistinguishable (see top panels of Figures 5 and
6).  Moreover the distributions in the top panels of Figures 2 and 3 (all-sky
searches) are indistinguishable from the corresponding distributions in Figures
5 and 6 (directed searches).  Since the signal model for all-sky searches and
the corresponding model for directed searches differ only in the number of the
unknown parameters in the phase of the signal this confirms an effective
decorrelation of the amplitude and the phase parameters.  This fact will enable
us (in Sec.\ 6 below) to give an assessment of the rms errors of the phase
parameters independently of the physical mechanisms generating the gravitational
waves from a spinning neutron star.

\begin{table}\begin{center}
\begin{tabular}{|c|c|c|c|c|c|c|c|c|c|c|}\hline
detector &&
$\frac{\sigma(h_o)}{h_o}$ &
$\sigma(\theta)$ (rad) &
$\Delta\Omega$ (sr) &
$\frac{\sigma(f_o)}{f_o}$ &
$\frac{\sigma({\fko 1})}{\vert{\fko 1}\vert}$ &
$\frac{\sigma({\fko 2})}{\vert{\fko 2}\vert}$ &
$\frac{\sigma({\fko 3})}{\vert{\fko 3}\vert}$ & 
$\frac{\sigma({\fko 4})}{\vert{\fko 4}\vert}$ & $d$ \\ 
\hline\hline
GEO600
& $q_{0.25}$
& $\scriptstyle4.3\times10^{-2}$
& $\scriptstyle3.5\times10^{-2}$
& $\scriptstyle1.1\times10^{-6}$
& $\scriptstyle5.3\times10^{-8}$
& $\scriptstyle6.4\times10^{-6}$
& $\scriptstyle1.7\times10^{-3}$
& $\scriptstyle6.4\times10^{-2}$
& $\scriptstyle8.5$
& $\scriptstyle17$
\\ \cline{2-11}wideband
& $q_{0.5}$
& $\scriptstyle9.0\times10^{-2}$
& $\scriptstyle6.6\times10^{-2}$
& $\scriptstyle3.6\times10^{-6}$
& $\scriptstyle9.8\times10^{-8}$
& $\scriptstyle1.0\times10^{-5}$
& $\scriptstyle3.1\times10^{-3}$
& $\scriptstyle1.0\times10^{-1}$
& $\scriptstyle16$
& $\scriptstyle28$
\\ \cline{2-11}noise
& $q_{0.75}$
& $\scriptstyle5.2\times10^{-1}$
& $\scriptstyle1.7\times10^{-1}$
& $\scriptstyle2.4\times10^{-5}$
& $\scriptstyle2.6\times10^{-7}$
& $\scriptstyle2.2\times10^{-5}$
& $\scriptstyle8.2\times10^{-3}$
& $\scriptstyle2.2\times10^{-1}$
& $\scriptstyle42$
& $\scriptstyle37$
\\ \hline\hline
GEO600
& $q_{0.25}$
& $\scriptstyle2.0\times10^{-1}$
& $\scriptstyle2.2\times10^{-1}$
& $\scriptstyle1.7\times10^{-8}$
& $\scriptstyle6.6\times10^{-9}$
& $\scriptstyle7.7\times10^{-7}$
& $\scriptstyle2.1\times10^{-4}$
& $\scriptstyle7.7\times10^{-3}$
& $\scriptstyle1.1$
& $\scriptstyle47$
\\ \cline{2-11}narrowband
& $q_{0.5}$
& $\scriptstyle5.3\times10^{-1}$
& $\scriptstyle2.6\times10^{-1}$
& $\scriptstyle7.1\times10^{-8}$ 
& $\scriptstyle1.4\times10^{-8}$
& $\scriptstyle1.3\times10^{-6}$
& $\scriptstyle4.5\times10^{-4}$
& $\scriptstyle1.3\times10^{-2}$
& $\scriptstyle2.3$
& $\scriptstyle160$
\\ \cline{2-11}noise
& $q_{0.75}$
& $\scriptstyle1.5$
& $\scriptstyle3.0\times10^{-1}$
& $\scriptstyle8.4\times10^{-7}$
& $\scriptstyle5.1\times10^{-8}$
& $\scriptstyle4.5\times10^{-6}$
& $\scriptstyle1.6\times10^{-3}$
& $\scriptstyle4.5\times10^{-2}$
& $\scriptstyle8.2$
& $\scriptstyle280$
\\ \hline\hline
initial
& $q_{0.25}$
& $\scriptstyle1.3\times10^{-2}$
& $\scriptstyle1.1\times10^{-2}$
& $\scriptstyle7.9\times10^{-8}$
& $\scriptstyle1.4\times10^{-8}$
& $\scriptstyle1.8\times10^{-6}$
& $\scriptstyle4.5\times10^{-4}$
& $\scriptstyle1.9\times10^{-2}$
& $\scriptstyle2.3$
& $\scriptstyle56$
\\ \cline{2-11}Hanford
& $q_{0.5}$
& $\scriptstyle2.8\times10^{-2}$
& $\scriptstyle2.0\times10^{-2}$
& $\scriptstyle2.6\times10^{-7}$
& $\scriptstyle2.6\times10^{-8}$
& $\scriptstyle2.9\times10^{-6}$
& $\scriptstyle8.1\times10^{-4}$
& $\scriptstyle2.9\times10^{-2}$
& $\scriptstyle4.1$
& $\scriptstyle96$
\\ \cline{2-11}
& $q_{0.75}$
& $\scriptstyle1.5\times10^{-1}$
& $\scriptstyle5.1\times10^{-2}$
& $\scriptstyle1.8\times10^{-6}$
& $\scriptstyle6.9\times10^{-8}$
& $\scriptstyle6.2\times10^{-6}$
& $\scriptstyle2.2\times10^{-3}$
& $\scriptstyle6.3\times10^{-2}$
& $\scriptstyle11$
& $\scriptstyle120$
\\ \hline\hline
advanced
& $q_{0.25}$
& $\scriptstyle1.3\times10^{-3}$
& $\scriptstyle1.1\times10^{-3}$
& $\scriptstyle7.9\times10^{-10}$
& $\scriptstyle1.4\times10^{-9}$
& $\scriptstyle1.8\times10^{-7}$
& $\scriptstyle4.5\times10^{-5}$
& $\scriptstyle1.8\times10^{-3}$
& $\scriptstyle0.23$
& $\scriptstyle560$
\\ \cline{2-11}Hanford
& $q_{0.5}$
& $\scriptstyle2.8\times10^{-3}$
& $\scriptstyle2.0\times10^{-3}$
& $\scriptstyle2.6\times10^{-9}$
& $\scriptstyle2.6\times10^{-9}$
& $\scriptstyle2.9\times10^{-7}$
& $\scriptstyle8.1\times10^{-5}$
& $\scriptstyle2.9\times10^{-3}$
& $\scriptstyle0.41$
& $\scriptstyle960$
\\ \cline{2-11}
& $q_{0.75}$
& $\scriptstyle1.5\times10^{-2}$
& $\scriptstyle5.0\times10^{-3}$
& $\scriptstyle1.7\times10^{-8}$
& $\scriptstyle6.9\times10^{-9}$
& $\scriptstyle6.2\times10^{-7}$
& $\scriptstyle2.2\times10^{-4}$
& $\scriptstyle6.3\times10^{-3}$
& $\scriptstyle1.1$
& $\scriptstyle1200$
\\ \hline\hline
initial
& $q_{0.25}$
& $\scriptstyle1.2\times10^{-2}$
& $\scriptstyle1.0\times10^{-2}$
& $\scriptstyle6.2\times10^{-8}$
& $\scriptstyle1.3\times10^{-8}$
& $\scriptstyle1.8\times10^{-6}$
& $\scriptstyle4.0\times10^{-4}$
& $\scriptstyle1.8\times10^{-2}$
& $\scriptstyle2.0$
& $\scriptstyle58$
\\ \cline{2-11}Livingston
& $q_{0.5}$
& $\scriptstyle2.7\times10^{-2}$
& $\scriptstyle2.0\times10^{-2}$
& $\scriptstyle1.8\times10^{-7}$ 
& $\scriptstyle2.2\times10^{-8}$
& $\scriptstyle2.7\times10^{-6}$
& $\scriptstyle7.0\times10^{-4}$
& $\scriptstyle2.7\times10^{-2}$
& $\scriptstyle3.6$
& $\scriptstyle97$
\\ \cline{2-11}
& $q_{0.75}$
& $\scriptstyle1.4\times10^{-1}$ 
& $\scriptstyle4.7\times10^{-2}$
& $\scriptstyle1.1\times10^{-6}$
& $\scriptstyle5.7\times10^{-8}$
& $\scriptstyle5.8\times10^{-6}$
& $\scriptstyle1.8\times10^{-3}$
& $\scriptstyle5.8\times10^{-2}$
& $\scriptstyle9.1$
& $\scriptstyle120$
\\ \hline\hline
advanced
& $q_{0.25}$
& $\scriptstyle1.2\times10^{-3}$
& $\scriptstyle1.0\times10^{-3}$
& $\scriptstyle6.2\times10^{-10}$
& $\scriptstyle1.3\times10^{-9}$
& $\scriptstyle1.8\times10^{-7}$
& $\scriptstyle4.0\times10^{-5}$
& $\scriptstyle1.8\times10^{-3}$
& $\scriptstyle0.20$
& $\scriptstyle580$
\\ \cline{2-11}Livingston
& $q_{0.5}$
& $\scriptstyle2.7\times10^{-3}$
& $\scriptstyle1.9\times10^{-3}$
& $\scriptstyle1.8\times10^{-9}$
& $\scriptstyle2.2\times10^{-9}$
& $\scriptstyle2.7\times10^{-7}$
& $\scriptstyle7.0\times10^{-5}$
& $\scriptstyle2.7\times10^{-3}$
& $\scriptstyle0.35$
& $\scriptstyle970$
\\ \cline{2-11}
& $q_{0.75}$
& $\scriptstyle1.4\times10^{-2}$
& $\scriptstyle4.7\times10^{-3}$
& $\scriptstyle1.1\times10^{-8}$
& $\scriptstyle5.7\times10^{-9}$
& $\scriptstyle5.7\times10^{-7}$
& $\scriptstyle1.8\times10^{-4}$
& $\scriptstyle5.8\times10^{-3}$
& $\scriptstyle0.91$
& $\scriptstyle1200$
\\ \hline\hline
& $q_{0.25}$
& $\scriptstyle1.1\times10^{-2}$
& $\scriptstyle9.6\times10^{-3}$
& $\scriptstyle5.6\times10^{-8}$
& $\scriptstyle1.2\times10^{-8}$
& $\scriptstyle1.5\times10^{-6}$
& $\scriptstyle3.8\times10^{-4}$
& $\scriptstyle1.6\times10^{-2}$
& $\scriptstyle1.9$
& $\scriptstyle57$
\\ \cline{2-11}VIRGO
& $q_{0.5}$
& $\scriptstyle2.5\times10^{-2}$
& $\scriptstyle1.8\times10^{-2}$
& $\scriptstyle1.8\times10^{-7}$
& $\scriptstyle2.2\times10^{-8}$
& $\scriptstyle2.4\times10^{-6}$
& $\scriptstyle6.9\times10^{-4}$ 
& $\scriptstyle2.5\times10^{-2}$ 
& $\scriptstyle3.5$
& $\scriptstyle100$
\\ \cline{2-11}
& $q_{0.75}$
& $\scriptstyle1.3\times10^{-1}$
& $\scriptstyle4.4\times10^{-2}$
& $\scriptstyle1.3\times10^{-6}$
& $\scriptstyle6.1\times10^{-8}$
& $\scriptstyle5.9\times10^{-6}$
& $\scriptstyle1.9\times10^{-3}$
& $\scriptstyle5.9\times10^{-2}$
& $\scriptstyle9.7$
& $\scriptstyle140$
\\ \hline\hline
& $q_{0.25}$
& $\scriptstyle1.2\times10^{-1}$
& $\scriptstyle1.2\times10^{-1}$
& $\scriptstyle4.6\times10^{-7}$
& $\scriptstyle3.6\times10^{-8}$
& $\scriptstyle4.7\times10^{-6}$
& $\scriptstyle1.1\times10^{-3}$
& $\scriptstyle4.7\times10^{-2}$
& $\scriptstyle5.7$
& $\scriptstyle17$
\\ \cline{2-11}TAMA300
& $q_{0.5}$
& $\scriptstyle3.0\times10^{-1}$
& $\scriptstyle1.5\times10^{-1}$
& $\scriptstyle1.8\times10^{-6}$
& $\scriptstyle7.4\times10^{-8}$
& $\scriptstyle7.8\times10^{-6}$
& $\scriptstyle2.3\times10^{-3}$
& $\scriptstyle7.9\times10^{-2}$
& $\scriptstyle12$
& $\scriptstyle33$
\\ \cline{2-11}
& $q_{0.75}$
& $\scriptstyle1.3$
& $\scriptstyle3.1\times10^{-1}$
& $\scriptstyle2.0\times10^{-5}$
& $\scriptstyle2.4\times10^{-7}$
& $\scriptstyle2.5\times10^{-5}$
& $\scriptstyle7.4\times10^{-3}$
& $\scriptstyle2.6\times10^{-1}$
& $\scriptstyle38$
& $\scriptstyle46$
\\ \hline
\end{tabular}\end{center}
\caption{The quartiles \cite{quartiles} for the Monte Carlo simulated
distributions of the rms errors of the signal parameters for the individual
detectors in the case of all-sky searches.  The observation time $T_o=120$ days.
The neutron star parameters are the same as in Figure 1, the frequency $f_o=500$
Hz, and the spindown age $\tau=40$ years.  The model of the signal's phase is
described by $s_1=4$, $s_2=3$, and $s_3=0$ (cf.\ Table 1). The dimensionless 
amplitude of the waveform is $h_o=1.1\times10^{-23}$. For the GEO600 detector we 
use two noise curves: wideband and narrowband tuned to 1 kHz with the bandwidth 
of 30 Hz. The last column shows the quartiles of the Monte Carlo simulated
distributions of the signal-to-noise ratio $d$ [given by Eq.\ (\ref{snr})], it 
is taken from Table II of Paper I.}\end{table}

\begin{table}[ht]\begin{center}
\begin{tabular}{|c|c|c|c|c|c|c|c|c|c|}\hline
detector &&
$\frac{\sigma(h_o)}{h_o}$ &
$\sigma(\theta)$ (rad) &
$\frac{\sigma(f_o)}{f_o}$ &
$\frac{\sigma({\fko 1})}{\vert{\fko 1}\vert}$ &
$\frac{\sigma(\stackrel{(2)}{f_o})}{\vert\stackrel{(2)}{ f_o}\vert}$ &
$\frac{\sigma(\stackrel{(3)}{f_o})}{\vert\stackrel{(3)}{ f_o}\vert}$ & 
$\frac{\sigma(\stackrel{(4)}{f_o})}{\vert\stackrel{(4)}{ f_o}\vert}$ & $d$ \\ 
\hline\hline
GEO600
& $q_{0.25}$
& $\scriptstyle4.3\times10^{-2}$
& $\scriptstyle3.4\times10^{-2}$
& $\scriptstyle6.7\times10^{-12}$
& $\scriptstyle5.0\times10^{-9}$
& $\scriptstyle4.7\times10^{-6}$
& $\scriptstyle6.7\times10^{-4}$
& $\scriptstyle0.40$
& $\scriptstyle17$
\\ \cline{2-10}wideband
& $q_{0.5}$
& $\scriptstyle8.8\times10^{-2}$
& $\scriptstyle6.6\times10^{-2}$
& $\scriptstyle1.1\times10^{-11}$
& $\scriptstyle8.0\times10^{-9}$
& $\scriptstyle7.4\times10^{-6}$
& $\scriptstyle1.1\times10^{-3}$
& $\scriptstyle0.64$
& $\scriptstyle28$
\\ \cline{2-10}noise
& $q_{0.75}$
& $\scriptstyle5.0\times10^{-1}$
& $\scriptstyle1.7\times10^{-1}$
& $\scriptstyle2.3\times10^{-11}$
& $\scriptstyle1.7\times10^{-8}$
& $\scriptstyle1.6\times10^{-5}$
& $\scriptstyle2.3\times10^{-3}$
& $\scriptstyle1.4$
& $\scriptstyle37$
\\ \hline\hline
GEO600
& $q_{0.25}$
& $\scriptstyle2.0\times10^{-1}$
& $\scriptstyle2.2\times10^{-1}$
& $\scriptstyle8.0\times10^{-13}$
& $\scriptstyle6.0\times10^{-10}$
& $\scriptstyle5.6\times10^{-7}$
& $\scriptstyle8.0\times10^{-5}$
& $\scriptstyle0.048$
& $\scriptstyle47$
\\ \cline{2-10}narrowband
& $q_{0.5}$
& $\scriptstyle5.3\times10^{-1}$
& $\scriptstyle2.6\times10^{-1}$
& $\scriptstyle1.4\times10^{-12}$
& $\scriptstyle1.0\times10^{-9}$
& $\scriptstyle9.6\times10^{-7}$
& $\scriptstyle1.4\times10^{-4}$
& $\scriptstyle0.083$
& $\scriptstyle160$
\\ \cline{2-10}noise
& $q_{0.75}$
& $\scriptstyle1.5$
& $\scriptstyle3.0\times10^{-1}$
& $\scriptstyle4.6\times10^{-12}$
& $\scriptstyle3.5\times10^{-9}$
& $\scriptstyle3.2\times10^{-6}$
& $\scriptstyle4.6\times10^{-4}$
& $\scriptstyle0.28$
& $\scriptstyle280$
\\ \hline\hline
initial
& $q_{0.25}$
& $\scriptstyle1.3\times10^{-2}$
& $\scriptstyle1.1\times10^{-2}$
& $\scriptstyle2.0\times10^{-12}$
& $\scriptstyle1.5\times10^{-9}$
& $\scriptstyle1.4\times10^{-6}$
& $\scriptstyle2.0\times10^{-4}$
& $\scriptstyle0.12$
& $\scriptstyle56$
\\ \cline{2-10}Hanford
& $q_{0.5}$
& $\scriptstyle2.8\times10^{-2}$
& $\scriptstyle2.0\times10^{-2}$
& $\scriptstyle3.1\times10^{-12}$
& $\scriptstyle2.4\times10^{-9}$
& $\scriptstyle2.2\times10^{-6}$
& $\scriptstyle3.1\times10^{-4}$
& $\scriptstyle0.19$
& $\scriptstyle96$
\\ \cline{2-10}
& $q_{0.75}$
& $\scriptstyle1.5\times10^{-1}$
& $\scriptstyle5.1\times10^{-2}$
& $\scriptstyle6.7\times10^{-12}$
& $\scriptstyle5.1\times10^{-9}$
& $\scriptstyle4.7\times10^{-6}$
& $\scriptstyle6.7\times10^{-4}$
& $\scriptstyle0.41$
& $\scriptstyle120$
\\ \hline\hline
advanced
& $q_{0.25}$
& $\scriptstyle1.3\times10^{-3}$
& $\scriptstyle1.1\times10^{-3}$
& $\scriptstyle2.0\times10^{-13}$
& $\scriptstyle1.5\times10^{-10}$
& $\scriptstyle1.4\times10^{-7}$
& $\scriptstyle2.0\times10^{-5}$
& $\scriptstyle0.012$
& $\scriptstyle560$
\\ \cline{2-10}Hanford
& $q_{0.5}$
& $\scriptstyle2.8\times10^{-3}$
& $\scriptstyle2.0\times10^{-3}$
& $\scriptstyle3.1\times10^{-13}$
& $\scriptstyle2.3\times10^{-10}$
& $\scriptstyle2.2\times10^{-7}$
& $\scriptstyle3.1\times10^{-5}$
& $\scriptstyle0.019$
& $\scriptstyle960$
\\ \cline{2-10}
& $q_{0.75}$
& $\scriptstyle1.5\times10^{-2}$
& $\scriptstyle5.1\times10^{-3}$
& $\scriptstyle6.7\times10^{-13}$
& $\scriptstyle5.0\times10^{-10}$
& $\scriptstyle4.7\times10^{-7}$
& $\scriptstyle6.7\times10^{-5}$
& $\scriptstyle0.040$
& $\scriptstyle1200$
\\ \hline\hline
initial
& $q_{0.25}$
& $\scriptstyle1.2\times10^{-2}$
& $\scriptstyle1.0\times10^{-2}$
& $\scriptstyle2.0\times10^{-12}$
& $\scriptstyle1.5\times10^{-9}$
& $\scriptstyle1.4\times10^{-6}$
& $\scriptstyle2.0\times10^{-4}$
& $\scriptstyle0.12$
& $\scriptstyle58$
\\ \cline{2-10}Livingston
& $q_{0.5}$
& $\scriptstyle2.6\times10^{-2}$
& $\scriptstyle2.0\times10^{-2}$
& $\scriptstyle3.0\times10^{-12}$
& $\scriptstyle2.3\times10^{-9}$
& $\scriptstyle2.1\times10^{-6}$
& $\scriptstyle3.0\times10^{-4}$
& $\scriptstyle0.18$
& $\scriptstyle97$
\\ \cline{2-10}
& $q_{0.75}$
& $\scriptstyle1.4\times10^{-2}$
& $\scriptstyle4.7\times10^{-2}$
& $\scriptstyle6.7\times10^{-12}$
& $\scriptstyle5.0\times10^{-9}$
& $\scriptstyle4.6\times10^{-6}$
& $\scriptstyle6.6\times10^{-4}$
& $\scriptstyle0.40$
& $\scriptstyle120$
\\ \hline\hline
advanced
& $q_{0.25}$
& $\scriptstyle1.2\times10^{-3}$
& $\scriptstyle1.0\times10^{-3}$
& $\scriptstyle2.0\times10^{-13}$
& $\scriptstyle1.5\times10^{-10}$
& $\scriptstyle1.4\times10^{-7}$
& $\scriptstyle2.0\times10^{-5}$
& $\scriptstyle0.012$
& $\scriptstyle580$
\\ \cline{2-10}Livingston
& $q_{0.5}$
& $\scriptstyle2.6\times10^{-3}$
& $\scriptstyle1.9\times10^{-3}$
& $\scriptstyle3.0\times10^{-13}$
& $\scriptstyle2.3\times10^{-10}$
& $\scriptstyle2.1\times10^{-7}$
& $\scriptstyle3.0\times10^{-5}$
& $\scriptstyle0.018$
& $\scriptstyle970$
\\ \cline{2-10}
& $q_{0.75}$
& $\scriptstyle1.4\times10^{-2}$
& $\scriptstyle4.7\times10^{-3}$
& $\scriptstyle6.6\times10^{-13}$
& $\scriptstyle5.0\times10^{-10}$
& $\scriptstyle4.6\times10^{-7}$
& $\scriptstyle6.6\times10^{-5}$
& $\scriptstyle0.040$
& $\scriptstyle1200$
\\ \hline\hline
& $q_{0.25}$
& $\scriptstyle1.1\times10^{-2}$
& $\scriptstyle9.5\times10^{-3}$
& $\scriptstyle1.7\times10^{-12}$
& $\scriptstyle1.3\times10^{-9}$
& $\scriptstyle1.2\times10^{-6}$
& $\scriptstyle1.7\times10^{-4}$
& $\scriptstyle0.10$
& $\scriptstyle57$
\\ \cline{2-10}VIRGO
& $q_{0.5}$
& $\scriptstyle2.5\times10^{-2}$
& $\scriptstyle1.8\times10^{-2}$
& $\scriptstyle2.6\times10^{-12}$
& $\scriptstyle2.0\times10^{-9}$
& $\scriptstyle1.8\times10^{-6}$ 
& $\scriptstyle2.6\times10^{-4}$ 
& $\scriptstyle0.16$
& $\scriptstyle100$
\\ \cline{2-10}
& $q_{0.75}$
& $\scriptstyle1.4\times10^{-1}$
& $\scriptstyle4.4\times10^{-2}$
& $\scriptstyle6.4\times10^{-12}$
& $\scriptstyle4.8\times10^{-9}$
& $\scriptstyle4.5\times10^{-6}$
& $\scriptstyle6.4\times10^{-4}$
& $\scriptstyle0.39$
& $\scriptstyle140$
\\ \hline\hline
& $q_{0.25}$
& $\scriptstyle1.2\times10^{-1}$
& $\scriptstyle1.2\times10^{-1}$
& $\scriptstyle5.2\times10^{-12}$
& $\scriptstyle3.9\times10^{-9}$
& $\scriptstyle3.6\times10^{-6}$
& $\scriptstyle5.2\times10^{-4}$
& $\scriptstyle0.31$
& $\scriptstyle17$
\\ \cline{2-10}TAMA300
& $q_{0.5}$
& $\scriptstyle3.0\times10^{-1}$
& $\scriptstyle1.5\times10^{-1}$
& $\scriptstyle8.8\times10^{-12}$
& $\scriptstyle6.6\times10^{-9}$
& $\scriptstyle6.1\times10^{-6}$
& $\scriptstyle8.8\times10^{-4}$
& $\scriptstyle0.53$
& $\scriptstyle33$
\\ \cline{2-10}
& $q_{0.75}$
& $\scriptstyle1.3$
& $\scriptstyle3.0\times10^{-1}$
& $\scriptstyle2.8\times10^{-11}$
& $\scriptstyle2.1\times10^{-8}$
& $\scriptstyle2.0\times10^{-5}$
& $\scriptstyle2.8\times10^{-3}$
& $\scriptstyle1.7$
& $\scriptstyle46$
\\ \hline
\end{tabular}\end{center}
\caption{The quartiles \cite{quartiles} for the Monte Carlo simulated
distributions of the rms errors of the signal parameters for the individual
detectors in the case when the position of the source in the sky is known.  The
observation time $T_o=120$ days.  The neutron star parameters are the same as in
Figure 1, the frequency $f_o=500$ Hz, and the spindown age $\tau=40$ years.  The
model of the signal's phase is described by $s_1=4$, $s_2=3$, and $s_3=0$ (cf.\
Table 1). The dimensionless amplitude of the waveform is 
$h_o=1.1\times10^{-23}$. For the GEO600 detector we use two noise curves:  
wideband and narrowband tuned to 1 kHz with the bandwidth of 30 Hz. The last 
column shows the quartiles of the Monte Carlo simulated distributions of the 
signal-to-noise ratio $d$ [given by Eq.\ (\ref{snr})], it is taken from Table 
II of Paper I.}\end{table}

\section{Estimation of the proper motion of the neutron star}

To describe the proper motion of the neutron star we shall introduce the 
parameters
that are customarily used by the pulsar astronomers.  When the star
moves with respect to the SSB reference frame its right ascension $\alpha$ and
declination $\delta$ change.  To first order we can assume that they change
linearly with time, i.e.\ we have
\bea
\alpha = \alpha_0 + \mu_{\alpha} t,\quad
\delta = \delta_0 + \mu_{\delta} t.
\eea
Let us denote by ${\bf v}_{\rm ns}$ the velocity of the star w.r.t.\ the SSB
reference frame and by ${\bf v}_{{\rm ns}\perp}$ the component of the velocity
${\bf v}_{\rm ns}$ perpendicular to the unit vector ${\bf n}_0$ in the direction
of the star (${\bf n}_0$ is along the line of sight to the star at $t=0$ for the
observer located in the SSB).  To first order in time it is easy to express the
velocity ${\bf v}_{{\rm ns}\perp}$ in terms of the parameters $\mu_{\alpha}$ and
$\mu_{\delta}$.  We have
\bea
{\bf v}_{{\rm ns}\perp} = r_o \left(\begin{array}{c}
-\mu_{\alpha}\sin\alpha_0\cos\delta_0 - \mu_{\delta}\cos\alpha_0\sin\delta_0\\
\mu_{\alpha}\cos\alpha_0\cos\delta_0 - \mu_{\delta}\sin\alpha_0\sin\delta_0\\ 
\mu_{\delta}\cos\delta_0 \end{array}\right).
\eea
The total proper motion of the star $\mu:=|{\bf v}_{{\rm ns}\perp}|/r_o$ is 
given in terms of $\mu_{\alpha}$ and $\mu_{\delta}$ as follows:
\be
\mu = \sqrt{\mu_{\delta}^2 + \mu_{\alpha}^2\cos^2\delta_0}.
\ee

In our Monte Carlo simulation we have considered the extreme case of a nearby
neutron star at a distance of 40 pc.  We have taken the transverse velocity 
$|{\bf v}_{{\rm ns}\perp}|$ of the star to be $10^3$ km/s.  Careful modelling of 
the
pulsar statistics has shown that the pulsar velocities extend to these values
\cite{Lor97,cat95}.  For $|{\bf v}_{{\rm ns}\perp}|=10^3$ km/s and $r_0=40$ pc
the contribution (\ref{pmphase}) to the phase due to the proper motion is
$\sim$4 cycles for frequency $f_o=500$ Hz (this value we have used in the 
simulation) and it decreases linearly with distance.

One can show that in the extreme case we consider here the phase model 
consistent with the 1/4 of a cycle criterion for 120 days of observation time 
and young neutron star with spindown age $\tau=40$ years reads [cf.\ Eqs.\ 
(\ref{ap200}) and (\ref{pmphase})]
\be
\label{pmphase1}
\Psi(t) = \Phi_0 + 2\pi \sum_{k=0}^{4}{\fko k}\frac{t^{k+1}}{(k+1)!}
+ \frac{2\pi}{c} {\bf n}_0\cdot{\bf r}_{\rm ES}(t)
\sum_{k=0}^{3}{\fko k}\frac{t^k}{k!}
+ \frac{2\pi}{c} \left( {\bf n}_0\cdot{\bf r}_{\rm E}(t) +
\frac{{\bf v}_{{\rm ns}\perp}\cdot{\bf r}_{\rm ES}(t)}{r_o}\,t \right)f_o.
\ee
For the phase given by Eq.\ (\ref{pmphase1}) the two-component 
gravitational-wave signal defined by Eqs.\ (\ref{s1})--(\ref{s4}) depends on the 
following 14 parameters: 
$h_o,\alpha_0,\mu_\alpha,\delta_0,\mu_\delta,\psi,\iota,\theta,\Phi_0,f_o,{\fko 
1},\ldots,{\fko 4}$.

In the Monte Carlo simulation for each set of randomly generated angles
$\{\alpha_0,\delta_0,\psi,\iota,\theta\}$ we have also randomly chosen the
direction of the star's motion on the sky keeping the total proper motion $\mu$
of the star fixed and equal to ($10^3$ km/s)/(40 pc) $\cong 5.3\times10^{3}$
mas/yr (mas/yr stands for milliarcseconds per year, units usually used in pulsar
astronomy).

The results of the numerical simulations are summarized in Figure 7.  We have
assumed Hanford LIGO detector in the advanced configuration and 120 days of
observation time.  The extreme case considered here gives interesting accuracies
of the estimation of the proper motion parameters resulting in the median of the
distributions of $\sigma(\mu_{\alpha})$ and $\sigma(\mu_{\delta})$ to be
$\sim2\times10^3$ mas/yr (i.e. 40\% rms error).  
As the distance to the source is small the estimation
of the fourth spindown parameter ${\fko 4}$ with a good accuracy of $\sim$10\%
is also possible.

\section{Simplified models of the gravitational-wave signal}

The phase of the gravitational-wave signal from a spinning neutron star changes
at the detector on the Earth over a characteristic time of less than 100 ms
whereas the amplitude of the signal changes over one day.  This means that the
detection of a long continuous signal requires an accurate model for its phase.
As shown in Appendix A even 1/4 of a cycle difference between the filter and the
signal can lead to loss of the signal-to-noise ratio by 10\%.

Accurate modelling of the amplitude is not so crucial.  Consequently the
information on the phase parameters in the amplitudes of the signal is much
smaller than in the phase.  This can clearly be seen from our simulations
presented in Sec.\ 3 which show that the phase parameters can be estimated much
more accurately than the amplitude parameters (see Figures 1 to 6).  As a result
the phase parameters effectively decorrelate from the amplitude parameters.
This decorrelation is shown in our numerical simulations (see discussion at the
end of Sec.\ 3 and Figures 2, 3, 5, 6, see also Figure 15).  Therefore we can
expect to obtain a model of a signal with constant amplitudes (i.e.  with no
amplitude modulation) that reproduces the covariance matrix for the phase
parameters of the full signal.  Such a model is presented in Sec.\ 5.1 below.

We also find it useful to have a {\em linear parametrization} of the phase of
the signal i.e.\ a set of parameters such that the phase is a linear function of
the parameters.  One advantage of the linear parametrization is that the maximum
likelihood estimators of the parameters are unbiased \cite{KKS}.  The other
property of a linearly parametrized phase is that the covariance matrix is
independent of the values of the parameters.  In real data analysis schemes
optimization of the codes when using a linear parametrization would be
independent of the values of the parameters, i.e.\ of the parameter region
searched.  We find that the linear parametrization is not possible in general.
In Sec.\ 5.2 we show however that it can be achieved approximately by neglecting
a term in the phase arising from Earth diurnal motion and all contributions from
the spindowns in the Earth spin and orbital terms.

In Sec.\ 5.3 we present another linear model (called linear model II) that has
been investigated by one of us \cite{K97a}.  It is obtained by neglecting
completely Earth diurnal motion and all spindowns in Earth orbital motion terms.
The simplest, polynomial in time phase model, is introduced in Sec.\ 5.4.  In
this model the motion of the detector w.r.t.\ the SSB is ignored.  In Sec.\ 5.5
we describe the results of Monte Carlo simulations of the covariance matrix to
compare the simplified models of the signal with the exact two-component model
given in Sec.\ 2.

\subsection{Constant amplitude model}

The constant amplitude model of the gravitational-wave signal has the phase of
the exact model whereas the time dependent amplitudes are replaced by some
constant effective values.  We have obtained these effective values from the
analytic formula for the signal-to-noise ratio given in Appendix B of Paper I.
The signal-to-noise ratio consists of a term proportional to the square root of
the observation time and some oscillatory term.  For the observation times
longer than several days the term proportional to the square root of the
observation time strongly dominates.  Our effective amplitudes take into account
only this dominant part of the signal-to-noise ratio. Thus for the 
two-component signal $h$ that we consider the constant amplitude model depends 
on the two constant amplitudes $h_{o1}$ and $h_{o2}$ 
as well as on the two initial phase parameters $\Phi_{01}$ and $\Phi_{02}$:
\be
\label{ac1a}
h(t) = h_{o1} \sin\left[\Phi(t)+\Phi_{01}\right] +
h_{o2} \sin\left[2\Phi(t)+\Phi_{02}\right],
\ee
where the phase $\Phi$ coincides with the exact phase model given by Eq.\ 
(\ref{ap200}). The constant amplitudes $h_{o1}$ and $h_{o2}$ read [cf.\ Eqs.\ 
(85), (86), and  Appendix B in Paper I]
\bea
\label{ac1b1}
h_{o1} &=& h_o\sin\zeta |\sin2\theta|
\sqrt{F_1(\iota)e_1(\delta)\cos4\psi + G_1(\iota)e_2(\delta)},\\
\label{ac1b2}
h_{o2} &=& h_o\sin\zeta \sin^2\theta
\sqrt{F_2(\iota)e_1(\delta)\cos4\psi + G_2(\iota)e_2(\delta)},
\eea
where
\ben
F_1(\iota)&=&-\frac{1}{16}\sin^4\iota,\quad
F_2(\iota)=\frac{1}{4}\sin^4\iota,\\
G_1(\iota)&=&\frac{1}{16}\sin^2\iota\left(1+\cos^2\iota\right),\quad
G_2(\iota)=\frac{1}{4}\left(1+6\cos^2\iota+\cos^4\iota\right),\\
e_1(\delta) &=& 4j_1 \cos^4\delta,\quad
e_2(\delta) = 4j_2 - j_3\cos2\delta + j_1\cos^22\delta,\\
j_1 &=& \frac{1}{256} \left(
4 - 20\cos^2\lambda + 35\sin^22\gamma\cos^4\lambda \right),\\
j_2 &=& \frac{1}{1024} \left(
68 - 20\cos^2\lambda - 13\sin^22\gamma\cos^4\lambda \right),\\
j_3 &=& \frac{1}{128} \left(
28 - 44\cos^2\lambda + 5\sin^22\gamma\cos^4\lambda \right).
\een

\subsection{Linear model I}

The signal $h'$ of the linear model I has two constant amplitudes given by Eqs.\ 
(\ref{ac1b1}) and (\ref{ac1b2}) so it can be written as
\be
\label{ac2a}
h'(t) = h_{o1} \sin\left[\Phi'(t)+\Phi_{01}\right] +
h_{o2} \sin\left[2\Phi'(t)+\Phi_{02}\right].
\ee

The phase $\Phi'$ of the signal $h'$ differs from the exact phase model $\Phi$ 
given
by Eq.\ (\ref{ap200}) by neglecting all spindowns in the phase modulation
due to the orbital motion of the Earth and by discarding the component of the
vector ${\bf r}_{\rm d}$ defining the position of the detector w.r.t.\ the SSB
which is perpendicular to the ecliptic (see Sec.2.2 of Paper I). The 
function $\Phi'$ is thus given by
\bea
\label{ac2b}
\Phi'(t)&=&2\pi\sum_{k=0}^{s}{\fko k}\frac{t^{k+1}}{(k+1)!}
\nonumber\\&&
+\frac{2\pi f_o}{c}\left\{
\left(\cos\ve\sin\alpha\cos\delta+\sin\ve\sin\delta\right)
\left[R_{ES}\sin\left(\phi_o+\Omega_o t\right)
+R_{E}\cos\lambda\cos\ve\sin\left(\phi_r+\Omega_r t\right)\right]
\right.\nonumber\\&&\left.
+\cos\alpha\cos\delta
\left[R_{ES}\cos\left(\phi_o+\Omega_o t\right)
+R_{E}\cos\lambda\cos\left(\phi_r+\Omega_r t\right)\right]\right\}.
\eea

After introducing the two new parameters
\be
\label{a1a2}
\alpha_1 := f_o 
\left(\cos\ve\sin\alpha\cos\delta+\sin\ve\sin\delta\right),\quad
\alpha_2 := f_o \cos\alpha\cos\delta,
\ee
the phase (\ref{ac2b}) becomes a linear function of the new parameters $f_o$, 
${\fko 1}$, \ldots,  ${\fko s}$, $\alpha_1$, and $\alpha_2$.

\subsection{Linear model II}

For the linear model II the signal $h''$ is defined by
\be
\label{ac3a}
h''(t) = h_{o1} \sin\left[\Phi''(t)+\Phi_{01}\right] +
h_{o2} \sin\left[2\Phi''(t)+\Phi_{02}\right],
\ee
where again the constant amplitudes $h_{o1}$ and $h_{o2}$ are given by Eqs.\ 
(\ref{ac1b1}) and (\ref{ac1b2}). In the phase $\Phi''$ of the signal compared 
with the exact phase model $\Phi$ given by Eq.\ (\ref{ap200}) we neglect all  
spindowns in the phase modulation due to the orbital motion of the Earth and we 
discard the whole modulation due to the Earth's diurnal motion. The function 
$\Phi''$ is thus given by
\bea
\label{ac3b}
\Phi''(t)&=&2\pi\sum_{k=0}^{s}{\fko k}\frac{t^{k+1}}{(k+1)!}
\nonumber\\&&
+\frac{2\pi f_o}{c}R_{ES}
\left[\cos\alpha\cos\delta\cos\left(\phi_o+\Omega_o t\right)
+\left(\cos\ve\sin\alpha\cos\delta+\sin\ve\sin\delta\right)
\sin\left(\phi_o+\Omega_o t\right)\right].
\eea

As in the previous model after introducing the parameters
\be
\alpha_1 := f_o 
\left(\cos\ve\sin\alpha\cos\delta+\sin\ve\sin\delta\right),\quad
\alpha_2 := f_o \cos\alpha\cos\delta,
\ee
the phase (\ref{ac3b}) becomes a linear function of the parameters $f_o$, 
${\fko 1}$, \ldots,  ${\fko s}$, $\alpha_1$, and $\alpha_2$.

A version of the above model including only the second component was used in a
simplified study of data analysis of gravitational-wave signals from spinning
neutron stars made by one of the authors \cite{K97a}.

Let us observe that both the constant amplitude model of Sec.\ 5.1 and the 
linear model I of Sec.\ 5.2 coincide with the linear model II for a detector 
located at the north or the south pole (then $\lambda=\pm90^\circ$).

\subsection{Polynomial model}

In this model the signal $h'''$ is given by
\be
\label{2poly1}
h'''(t) = h_{o1} \sin\left[\Phi'''(t)+\Phi_{01}\right] +
h_{o2} \sin\left[2\Phi'''(t)+\Phi_{02}\right],
\ee
with the constant amplitudes $h_{o1}$ and $h_{o2}$ defined by Eqs.\ 
(\ref{ac1b1}) and (\ref{ac1b2}). In the phase $\Phi'''$ of the signal 
(\ref{2poly1}) we discard all the terms due to the motion of the detector 
relative to the SSB. This leads to the following polynomial in time model of the 
phase
\be
\label{2poly2}
\Phi'''(t) = 2\pi\sum_{k=0}^{s}{\fko k}\frac{t^{k+1}}{(k+1)!}.
\ee

For the case of directed searches i.e.\ when the position of the source in the
sky is known, the phase is a linear function of the unknown parameters:  initial
phase, frequency, and spindowns and thus the exact phase model is linear.  In
Sec.\ 5.5 below we find that for directed searches the polynomial model
reproduces accurately the covariance matrix for the exact model.  We investigate
the polynomial phase model in more detail in Appendix C.

\subsection{Simplified models vs.\ the exact model}

We have carried out the Monte Carlo simulations to compare our simplified signal
models with the exact model given in Sec.\ 2.  In our comparisons we have
considered both all-sky searches and directed searches and we have chosen two
observation times:  7 days and 120 days.  We have found that the constant
amplitude model of Sec.\ 5.1 reproduces very accurately the results of the exact
model (see Figures \ref{f:comp7} and \ref{f:comp120} for results of Monte Carlo
simulations).  Also the linear model I reproduces the rms errors of the exact
model well (see Figures \ref{f:comp7} and \ref{f:comp120}).  We have compared
the percentiles of the distributions of the rms errors of the phase parameters
for this model with the percentiles of the distributions of the exact model.  We
have found that for all-sky searches and observation time of 7 days, frequency
$f_o = 500$ Hz, and spindown age $\tau = 40$ years, the maximum differences for
various parameters range from $\sim$15\% to $\sim$60\%.  For observation time of
120 days and the same frequency and spindown age the corresponding differences
range from $\sim$7\% to $\sim$50\%.  For the linear model II the deviations of
the rms errors from the exact model are large (see Figures \ref{f:comp7} and
\ref{f:comp120}).

We have found that for directed searches the polynomial phase model introduced
in Sec.\ 5.4 reproduces the results of the exact model very well.  For
observation time of 7 days the maximum percentile differences of the
distributions of the phase parameter errors are around 5\% and for observation
time of 120 days they are less then 1\%.

Numerical calculations of the covariance matrices for the constant amplitude
models are enormously simplified compared to the amplitude modulated signals.
Since the model of the phase is independent of the physical mechanisms
generating the gravitational radiation, the simpler constant amplitude models
characterize well a general continuous gravitational-wave signal and they can be
a useful tool in theoretical analysis of the gravitational-wave signals from
spinning neutron stars.  However we cannot compromise the detectability of
signals for simplicity of the constant amplitude and linear phase models and we
do not insist on using them in real data analysis schemes.  Nonetheless for
searches over certain limited parameter space or for shorter observation times
the simplified models (both constant amplitude and linear phase models) may be 
useful.

\section{Dependence of the covariance matrix on the observational parameters}

It is clear that covariance matrices of various signals considered in the
present paper depend on the observation time $T_o$.  We also notice that the
phases of our signals and consequently the covariance matrices depend also on
the deterministic phases $\phi_o$, $\phi_r$ and on the latitude $\lambda$ of the
detector's location [see Eqs.\ (\ref{ap200})--(\ref{n0res})].  The values of the
phases $\phi_o$ and $\phi_r$ depend on the initial time of the observation.  It
is important to know these dependencies and therefore we study them in detail in
this section.

The results of the previous section indicate that the constant amplitude linear
phase model of Sec.\ 5.2 reproduces well the covariance matrices of the exact
model.  For directed searches even the simpler polynomial model of Sec.\ 5.4
is adequate.  Therefore we adopt these two models to study the dependence of
covariance matrices on $T_o$, $\phi_o$, $\phi_r$, and $\lambda$.  Moreover we
restrict ourselves to the one-component versions of these models.

Both models of Sec.\ 5.2 and of Sec.\ 5.4 are the constant amplitude models. We 
first derive the general formulae for the covariance matrix valid for the 
constant 
amplitude signal with any phase model, provided the signal is narrowband around 
some frequency $f_o$. Such a one-component signal can be written as
\be
\label{ca1}
h(t;h_o,\boldsymbol{\zeta}) = h_o \sin\Psi(t;\boldsymbol{\zeta}),
\ee
where $h_o$ is a constant amplitude and $\boldsymbol{\zeta}$ denotes parameters 
entering the phase $\Psi$ of the signal. We collect all the signal parameters 
into the vector $\boldsymbol{\theta}=(h_o,\boldsymbol{\zeta})$.
Assuming that over the bandwidth of the signal the spectral density $S_h$ of the 
noise is nearly constant and equal to $S_h(f_o)$ and that during the observation 
time the phase $\Psi$ has many cycles the optimal signal-to-noise ratio 
$d:=\sqrt{(h|h)}$ for the signal (\ref{ca1}) equals
\be
\label{ca2}
d \cong \frac{h_o\sqrt{T_o}}{\sqrt{S_h(f_o)}}.
\ee
The Fisher matrix $\Gamma$ for that signal can be written as [cf.\ Eq.\ 
(\ref{gammaij})]:
\be
\label{ca3}
\Gamma_{\theta_i\theta_j} \cong \frac{2}{S_h(f_o)} \int_{-T_o/2}^{T_o/2}
\frac{\pa h}{\pa \theta_i}\frac{\pa h}{\pa \theta_j}\,dt,
\ee
where $T_o$ is the observation time and the observation interval is 
$\left[-T_o/2,T_o/2\right]$. The components of the matrix $\Gamma$ read:
\bea
\label{ca4a}
\Gamma_{h_oh_o} &\cong& \frac{d^2}{h_o^2},\\
\label{ca4b}
\Gamma_{h_o\zeta_i} &\cong& 0,\\
\label{ca4c}
\Gamma_{\zeta_i\zeta_j} &\cong& d^2 {\widetilde \Gamma}_{\zeta_i\zeta_j},
\eea
where the reduced Fisher matrix ${\widetilde \Gamma}$ is defined by
\be
\label{ca5}
{\widetilde \Gamma}_{\zeta_i\zeta_j} := \frac{1}{T_o}
\int_{-T_o/2}^{T_o/2} \frac{\pa\Psi}{\pa\zeta_i}\frac{\pa\Psi}{\pa\zeta_j}\,dt.
\ee

The covariance  matrix $C$ approximated by the inverse of the Fisher matrix 
$\Gamma$ has the components
\bea
\label{ca6a}
C_{h_oh_o} &\cong& \frac{h_o^2}{d^2},\\
\label{ca6b}
C_{h_o\zeta_i} &\cong& 0,\\
\label{ca6c}
C_{\zeta_i\zeta_j} &\cong& \frac{1}{d^2}
\left({\widetilde \Gamma}^{-1}\right)_{\zeta_i\zeta_j}.
\eea

It is sometimes convenient to replace the spindowns ${\fko k}$ by the 
dimensionless parameters $\omega_k$ defined as
\be
\label{omega}
\omega_k := \frac{2\pi}{(k+1)!} {\fko k} T_o^{k+1},
\ee
where $T_o$ is the observation time. The rms errors of the spindown parameters 
${\fko k}$ are related to the rms errors of the dimensionless parameters 
$\omega_k$ by
\be
\label{fko-omega}
\label{sigma}
\sigma(\fko k) = \frac{(k+1)!}{2\pi}\frac{1}{T_o^{k+1}}\sigma(\omega_k),\quad
k=0,\ldots,s.
\ee
The significance of the parameters $\omega_k$ can be explained as follows. In 
the polynomial phase model (discussed in Appendix C) the covariance matrix for 
the parameters $\omega_k$ is independent of the observation time $T_o$, so it is 
completely determined by the number of spindowns included in the phase. In all 
more complicated phase models considered in this paper, including the exact 
phase model introduced in Sec.\ 2, the phase is the sum of the polynomial in 
time part and extra terms due to the motion of the detector w.r.t.\ the SSB. 
The polynomial part gives dominant contribution to the number of cycles in the 
phase and it roughly determines how the covariance matrices depend on the  
observation time $T_o$. Using the parameters $\omega_k$ instead of ${\fko k}$ we 
absorb the polynomial phase model dependence on $T_o$ into the very definition 
of the parameters $\omega_k$. If the covariance matrix for the parameters 
$\omega_k$ in a non-polynomial phase model depends on $T_o$, this 
dependence is a measure of how this model is different from the polynomial phase 
model.

For all-sky searches we have studied the dependence of covariance matrices on
$T_o$, $\phi_r$, $\phi_o$, and $\lambda$ using the constant amplitude linear
phase signal of Sec.\ 5.2.  We have rewritten the phase of this signal in 
terms of the parameters (\ref{a1a2}) and (\ref{omega}) (then the phase
is a linear function of these parameters).  The signal can be written as [cf.\ 
Eq.\ (\ref{ac2b})]
\bea
\label{sig6a}
h(t;h_o,\boldsymbol{\zeta}) &=& h_o \sin\Psi(t;\boldsymbol{\zeta}),
\\
\label{sig6b}
\Psi(t;\boldsymbol{\zeta}) &=& \Phi_0 + 
\sum_{k=0}^{s}\omega_k\left(\frac{t}{T_o}\right)^{k+1}
+\frac{2\pi}{c}\left\{
\alpha_1\left[R_{ES}\sin\left(\phi_o+\Omega_o t\right)
+R_{E}\cos\lambda\cos\ve\sin\left(\phi_r+\Omega_r t\right)\right]
\right.\nonumber\\&&\left.
+\alpha_2
\left[R_{ES}\cos\left(\phi_o+\Omega_o t\right)
+R_{E}\cos\lambda\cos\left(\phi_r+\Omega_r t\right)\right]\right\},
\eea
where $\boldsymbol{\zeta} = 
(h_o,\Phi_0,\alpha_1,\alpha_2,\omega_0,\ldots,\omega_s)$.
Thus the signal (\ref{sig6a}) depends on $5+s$ parameters. We have computed the 
covariance matrices for this signal by means of Eqs.\ 
(\ref{ca6a})--(\ref{ca6c}). The results are discussed below.

To study the dependence of the rms errors of the phase parameters on the
observation time $T_o$ we have considered a wide range of observation times from
1 hour to 4000 days.  We have taken the latitude $\lambda$ of the LIGO detector
in Hanford and we have assumed the signal-to-noise ratio $d=10$.  We have
considered five phase models given by Eq.\ (\ref{sig6b}) with $s=0,\ldots,4$
spindowns.  The results of the numerical calculations are shown in Figure
\ref{f:liot}.  We see that for observation times $T_o$ up to $\sim$20 days the
rms parameter errors decrease.  For observation times from $\sim$20 days to
$\sim$100 days the rms errors for initial phase $\Phi_0$ and parameters
$\alpha_1$ and $\alpha_2$ tend to decrease.  The rate of the decrease is the
smaller the more spindown parameters enter the phase of the signal and for
higher spindown models the errors stay even constant for certain ranges of the
observation time, whereas the frequency and the spindown errors in these ranges
tend to increase.  For high spindowns there is a range of observation time for
which rms errors stay constant.  For observation times from $\sim$100 days to
$\sim$1000 days all the rms errors decrease and for observation times greater
than $\sim$1000 days the rms errors level out to constant values.  We have
verified that all these constant values are to a very good accuracy equal to the
rms errors of the polynomial phase model for the corresponding number of
spindown parameters (these constant values can be found in Table 4 of
Appendix C).  Thus for observation times of more than $\sim$3 years the effect
of the motion of the detector relative to the SSB has negligible effect for the
rms errors of the initial phase, the frequency, and the spindown parameters.

The dependence of the rms phase parameters errors on the phase $\phi_r$ which is
determined by the initial position of the Earth in its diurnal motion is small.
The errors vary by no more than $\sim$10\% for the observation time $T_o$ = 7
days and for the phase models (\ref{sig6b}) with $s=0,\ldots,4$ spindowns (see
Figure \ref{f:lirot}, where the model with $s=4$ spindowns is studied) assuming
the signal-to-noise ratio $d=10$.  This dependence weakens when the observation
time $T_o$ increases.

The variations of the rms errors of the phase parameters with the initial phase
$\phi_o$ of the Earth's orbital motion are larger and depend in a complicated
way on the observation time $T_o$ and the number $s$ of spindowns included in
the phase model.  For the observation time $T_o$ = 120 days and the phase model
(\ref{sig6b}) with $s=4$ spindowns (assuming the signal-to-noise ratio $d=10$)
the variations are less than $\sim$10\% for all parameters except for parameters
$\alpha_1$ and $\alpha_2$ where they are of order 100\% (see Figure
\ref{f:liorb}).  For the observation time $T_o$ = 120 days and the
signal-to-noise ratio $d=10$ the phase models (\ref{sig6b}) with $s=0,\ldots,3$
have the rms errors of all parameters that vary less with $\phi_o$ compared to
the $s=4$ model, with the exception of parameters $\alpha_1$ and $\alpha_2$, for
which the variations range from $\sim$500\% to $\sim$900\%.  For the observation
time $T_o$ = 7 days and the phase models (\ref{sig6b}) with $s=0,\ldots,4$
spindowns (assuming the signal-to-noise ratio $d=10$) the variations are less
than $\sim$10\% for all the parameters except again for the parameters 
$\alpha_1$ and
$\alpha_2$.  For these parameters the variations range from $\sim$$10^{-7}$\%
for $s=4$ up to $\sim$$10^{3}$\% for $s=0$.

The rms errors of the phase parameters exhibit a sharp increase when the
detector's location approaches the south or north pole.  The dependence is
weaker for the phase models (\ref{sig6b}) with smaller number $s$ of spindowns
included and it gets stronger when the observation time $T_o$ increases.
However the latitudes of the laser-interferometric detectors currently under
construction range from 30.56 to 52.25 degrees (see Table I in Paper I) and in
this interval the rms errors vary by no more than $\sim$30\% for the observation
time $T_o$ = 120 days and the phase model (\ref{sig6b}) with $s=4$ spindowns
(see Figure \ref{f:lilat}) assuming the signal-to-noise ratio $d=10$. The sharp 
increase of the rms errors when $\lambda$ tends to $\pm90^{\circ}$ explains why 
the linear phase model II of Sec.\ 5.3 gives the errors significantly larger 
compared to another constant amplitude models (see Figures 8 and 9).

In Table 4 we have collected the rms errors of the initial phase $\Phi_0$, the
parameters $\alpha_1$, $\alpha_2$, and the spindowns ${\fko k}$ ($k=0,\ldots,4$)
in the case of all-sky searches.  We have considered five phase models
(\ref{sig6b}) with $s=0,\ldots,4$ spindowns included and the observation times
$T_o$ of 7 and 120 days.  The errors are calculated for the signal-to-noise
ratio $d = 10$ and they are inversely proportional to $d$.  They are
independent of the values of the spindown parameters in the signal.  As a
reference we quote the values of the spindowns for the Crab pulsar:  ${\fko 1}
= -3.773 \times 10^{-10}$ s$^{-2}$, ${\fko 2} = 0.976 \times 10^{-20}$
s$^{-3}$, and ${\fko 3} = -0.615\times 10^{-30}$ s$^{-4}$.  We see that in
many cases the value of the rms error of a certain parameter does not change if
we go from a model with $k$ spindowns to a model with $k+1$ spindowns.  For
example for observation time of 7 days the rms error of frequency is almost the
same for all the models.  This merging of the errors is visible clearly in
Figure 10 and occurs for certain ranges of the observation time.  It is related
to the merging for the polynomial phase model studied in Appendix C (see Figures
14 and 15).

The accuracy $\Delta\Omega$ of the position of the source in the sky can be
expressed in terms of the rms errors of $\alpha_1$, $\alpha_2$, and $f_o$ by
means of the rule of propagation of errors.

We found in the previous section that for directed searches the polynomial phase
model is adequate.  Consequently in the case of directed searches we can
completely neglect the dependence of the covariance matrices on the initial
phases $\phi_r$ and $\phi_o$ and also on the latitude $\lambda$ of the
detector's location.  The rms errors of the initial phase $\Phi_0$ and the
dimensionless spindown parameters $\omega_k$ ($k=0,\ldots,4$) can be found in
Table 4 of Appendix C.  They are constant numbers, independent on the
observation time $T_o$.  However the rms errors of the spindown parameters
${\fko k}$ ($k=0,\ldots,4$) do depend on the observation time $T_o$ and this
dependence is described by Eq.\ (\ref{fko-omega}).

In Table 5 we have given the rms errors of the initial phase $\Phi_0$ and the
spindowns ${\fko k}$ ($k=0,\ldots,4$) for directed searches.  We have considered
five polynomial phase models with $s=0,\ldots,4$ spindowns included and the
observation times $T_o$ of 7 and 120 days.  The errors are calculated for the
signal-to-noise ratio $d = 10$ and they are inversely proportional to $d$.

The fact that in some cases (both in all-sky and directed searches) the errors
of the parameters do not change when we go to a model with one more spindown is
a result of a special choice of the initial time of the observation which was
chosen to be in the middle of the observation interval.  This effect is studied
in detail in Appendix C (see Figure 14).

\begin{table}[ht]\begin{center}
\begin{tabular}{|c|c|c|c|c|c|c|c|c|}\hline
$s$ &
\begin{tabular}{c}$\sigma(\Phi_0)$\\ (rad)\end{tabular} &
\begin{tabular}{c}$\sigma(\alpha_1)$\\ (s$^{-1}$)\end{tabular} &
\begin{tabular}{c}$\sigma(\alpha_2)$\\ (s$^{-1}$)\end{tabular} &
\begin{tabular}{c}$\sigma(f_o)$\\ (s$^{-1}$)\end{tabular} &
\begin{tabular}{c}$\sigma({\fko 1})$\\ (s$^{-2}$)\end{tabular} &
\begin{tabular}{c}$\sigma({\fko 2})$\\ (s$^{-3}$)\end{tabular} &
\begin{tabular}{c}$\sigma({\fko 3})$\\ (s$^{-4}$)\end{tabular} & 
\begin{tabular}{c}$\sigma({\fko 4})$\\ (s$^{-5}$)\end{tabular} \\ \hline
\multicolumn{9}{|c|}{$T_o$ = 7 days} \\ \hline
0 &
$1.8\times10^2$ &
1.6 &
0.21 &
$1.6\times10^{-4}$ &&&& \\ \hline
1 &
$4.9\times10^3$ &
1.6 &
1.6 &
$1.6\times10^{-4}$ &
$3.1\times10^{-11}$ &&& \\ \hline
2 &
$4.9\times10^3$ &
1.7 &
1.6 &
$1.7\times10^{-4}$ &
$3.1\times10^{-11}$ &
$2.4\times10^{-17}$ && \\ \hline
3 &
$4.9\times10^3$ &
1.7 &
1.6 &
$1.7\times10^{-4}$ &
$3.1\times10^{-11}$ &
$2.4\times10^{-17}$ &
$6.0\times10^{-22}$ & \\ \hline
4 &
$5.0\times10^3$ &
1.7 &
1.6 &
$1.7\times10^{-4}$ &
$3.2\times10^{-11}$ &
$1.0\times10^{-16}$ &
$6.0\times10^{-22}$ &
$2.0\times10^{-26}$ \\ \hline
\multicolumn{9}{|c|}{$T_o$ = 120 days} \\ \hline
0 &
0.57 &
$1.2\times10^{-3}$ &
$2.6\times10^{-4}$ &
$1.1\times10^{-7}$ &&&& \\ \hline
1 &
29 &
$1.7\times10^{-3}$ &
$9.2\times10^{-3}$ &
$1.1\times10^{-7}$ &
$1.7\times10^{-13}$ &&& \\ \hline
2 &
29 &
$8.8\times10^{-2}$ &
$1.4\times10^{-2}$ &
$8.8\times10^{-6}$ &
$1.7\times10^{-13}$ &
$3.3\times10^{-19}$ && \\ \hline
3 &
$2.7\times10^3$ &
0.14 &
0.85 &
$8.8\times10^{-6}$ &
$1.7\times10^{-11}$ &
$3.3\times10^{-19}$ &
$6.4\times10^{-25}$ & \\ \hline
4 &
$2.7\times10^3$ &
1.7 &
0.87 &
$1.6\times10^{-4}$ &
$1.7\times10^{-11}$ &
$6.5\times10^{-18}$ &
$6.4\times10^{-25}$ &
$2.5\times10^{-31}$ \\ \hline
\end{tabular}\end{center}
\caption{The rms errors of the initial phase $\Phi_0$, the parameters 
$\alpha_1$, $\alpha_2$, and the spindowns ${\fko k}$ ($k=0,\ldots,4$) for 
all-sky searches. We have approximated the gravitational-wave signal by the 
one-component constant amplitude and linear phase model of Sec.\ 5.2. We have 
considered five phase models with $s=0,\ldots,4$ spindowns included and the 
observation times $T_o$ of 7 and 120 days. The signal-to-noise ratio $d=10$. We 
have assumed the latitude $\lambda$ of the Hanford LIGO detector, we have also 
put $\phi_r=1.456$ and $\phi_o=0.123$.}\end{table}

\begin{table}[ht]\begin{center}
\begin{tabular}{|c|c|c|c|c|c|c|}\hline
$s$ &
\begin{tabular}{c}$\sigma(\Phi_0)$\\ (rad)\end{tabular} &
\begin{tabular}{c}$\sigma(f_o)$\\ (s$^{-1}$)\end{tabular} &
\begin{tabular}{c}$\sigma({\fko 1})$\\ (s$^{-2}$)\end{tabular} &
\begin{tabular}{c}$\sigma({\fko 2})$\\ (s$^{-3}$)\end{tabular} &
\begin{tabular}{c}$\sigma({\fko 3})$\\ (s$^{-4}$)\end{tabular} & 
\begin{tabular}{c}$\sigma({\fko 4})$\\ (s$^{-5}$)\end{tabular} \\ \hline
\multicolumn{7}{|c|}{$T_o$ = 7 days} \\ \hline
0 &
0.10 &
$9.1\times10^{-8}$ &&&& \\ \hline
1 &
0.15 &
$9.1\times10^{-8}$ &
$1.2\times10^{-12}$ &&& \\ \hline
2 &
0.15 &
$2.3\times10^{-7}$ &
$1.2\times10^{-12}$ &
$2.3\times10^{-17}$ && \\ \hline
3 &
0.19 &
$2.3\times10^{-7}$ &
$4.1\times10^{-12}$ &
$2.3\times10^{-17}$ &
$6.0\times10^{-22}$ & \\ \hline
4 &
0.19 &
$4.0\times10^{-7}$ &
$4.1\times10^{-12}$ &
$1.0\times10^{-16}$ &
$6.0\times10^{-22}$ &
$2.0\times10^{-26}$ \\ \hline
\multicolumn{7}{|c|}{$T_o$ = 120 days} \\ \hline
0 &
0.10 &
$5.3\times10^{-9}$ &&&& \\ \hline
1 &
0.15 &
$5.3\times10^{-9}$ &
$4.0\times10^{-15}$ &&& \\ \hline
2 &
0.15 &
$1.3\times10^{-8}$ &
$4.0\times10^{-15}$ &
$4.5\times10^{-21}$ && \\ \hline
3 &
0.19 &
$1.3\times10^{-8}$ &
$1.4\times10^{-14}$ &
$4.5\times10^{-21}$ &
$6.9\times10^{-27}$ & \\ \hline
4 &
0.19 &
$2.3\times10^{-8}$ &
$1.4\times10^{-14}$ &
$2.0\times10^{-20}$ &
$6.9\times10^{-27}$ &
$1.3\times10^{-32}$ \\ \hline
\end{tabular}\end{center}
\caption{The rms errors of the initial phase $\Phi_0$ and the spindowns ${\fko 
k}$ ($k=0,\ldots,4$) for directed searches. We have approximated the 
gravitational-wave signal by the one-component constant amplitude and polynomial 
phase model of Sec.\ 5.4. We have considered five phase models with 
$s=0,\ldots,4$ spindowns included and the observation times $T_o$ of 7 and 120 
days. We have assumed the signal-to-noise ratio $d=10$.}\end{table}

\section*{Acknowledgments}

We would like to thank the Albert Einstein Institute, Max Planck Institute for
Gravitational Physics for hospitality.  We would also like to thank Bernard F.\
Schutz for many useful discussions, comments, and suggestions that led to a
considerable improvement of this work.  This work was supported in part by
Polish Science Committee grant KBN 2 P303D 021 11.

\appendix

\section{1/4 of a cycle criterion}

The fitting factor FF between a signal $h=h(t;\boldsymbol{\theta})$ and a filter 
 $h'=h'(t;\boldsymbol{\theta}')$ ($\boldsymbol{\theta}$ and 
$\boldsymbol{\theta}'$ are the parameters of the signal and the filter, 
respectively) is defined as \cite{A1}
\be
\label{apa1}
\text{FF} := \max_{\boldsymbol{\theta}'}
\frac{\left(h(t;\boldsymbol{\theta})\vert h'(t;\boldsymbol{\theta}')\right)}
{\sqrt{\left(h(t;\boldsymbol{\theta})\vert h(t;\boldsymbol{\theta})\right)}
\sqrt{\left(h'(t;\boldsymbol{\theta}')\vert h'(t;\boldsymbol{\theta}')\right)}}.
\ee
$1-\text{FF}$ gives the fraction of the signal-to-noise lost when using a filter
not perfectly matched to a signal. For narrowband signals around the frequency 
$f_o$ the scalar product $(\cdot\vert\cdot)$ can be computed from the formula
\be
\label{apa2}
(h_1\vert h_2)\cong \frac{2}{S_h(f_o)}\int_{-T_o/2}^{T_o/2}h_1(t)h_2(t)\,dt,
\ee
where $S_h$ is the one-sided noise spectral density and $T_o$ is the observation 
time.

Let us assume that the signal and the filter can be written as
\bea
\label{apa3a}
h(t;\boldsymbol{\theta}) &=& h_o \sin\Psi(t;\boldsymbol{\zeta}),\\
\label{apa3b}
h'(t;\boldsymbol{\theta}') &=& h'_o \sin\Psi'(t;\boldsymbol{\zeta}'),
\eea
where $h_o$ and $h'_o$ are constant amplitudes, $\boldsymbol{\zeta}$ and 
$\boldsymbol{\zeta}'$ denote the parameters entering the phases $\Psi$ and 
$\Psi'$ of the signal and the filter, respectively. We substitute Eqs.\ 
(\ref{apa3a}) and (\ref{apa3b}) into Eq.\ (\ref{apa1}). Using Eq.\ (\ref{apa2}) 
we obtain
\be
\label{apa4}
\text{FF} \cong \max_{\boldsymbol{\zeta}'} \frac{1}{T_o} \int_{-T_o/2}^{T_o/2} 
\cos\left[\Psi(t;\boldsymbol{\zeta})-\Psi'(t;\boldsymbol{\zeta}')\right]\,dt.
\ee

The fitting factor attains its maximum value of 1 when the functions $\Psi$ and
$\Psi'$ are the same and when values of the parameters in the signal and the
filter coincide.  When $\Psi$ and $\Psi'$ are not the same because in the phase
of the filter we have not taken into account some effects present in the signal
the fitting factor is less than 1.  Moreover the values of the parameter in the
filter that maximize the correlation integral will be biased, i.e.\ shifted away
from the true values in the signal.

For the simplest nontrivial case of the difference between the phase of the
filter and the phase of the signal consisting of a constant term (constant phase
$\phi$) and a term linear in time (with a constant frequency $f$) the fitting
factor (\ref{apa4}) equals
\be
\label{apa5}
\text{FF} \cong \max_{\phi} \frac{1}{T_o} \int_{-T_o/2}^{T_o/2}
\cos\left(2\pi ft + \phi\right)\,dt.
\ee
We easily get
\be
\label{apa6}
\text{FF} \cong \max_{\phi}
\left[\frac{\sin(\pi fT_o)}{\pi fT_o}\cos\phi\right]
= \frac{\sin(\pi fT_o)}{\pi fT_o}.
\ee
If the frequency $f$ is such that it produces not more than 1/4 of a cycle 
during the observation time $T_o$ (i.e.\ $fT_o\le 1/4$) then from Eq.\ 
(\ref{apa6}) we obtain
$$
\text{FF} \gtrsim \frac{\sin\frac{\pi}{4}}{\frac{\pi}{4}} \cong 0.900.
$$
Thus characterization of the 1/4 of a cycle criterion is the following:
{\em for the simplest nontrivial case discarding a term in the phase 
that contributes less than 1/4 of a cycle over observation time leads
to a loss in signal-to-noise ratio of not more than 10\%}.

\section{Number of cycles}

The model of the phase of the signal at the detector 
given by Eq.\ (\ref{ap200}) can be rewritten as
\be
\label{ap13}
\Psi = \Phi_0
+2\pi\sum_{k=0}^{s_1} N_k
+2\pi\sum_{k=0}^{s_2} N_k^{(o)}+2\pi\sum_{k=0}^{s_3} N_k^{(r)},
\ee
where $N_k$, $N_k^{(o)}$ and $N_k^{(r)}$ denote the numbers of cycles 
arising from the term polynomial in time, the orbital motion and the rotational 
motion of the detector, respectively:
\be
\label{ap15a}
N_k := {\fkg k}\frac{t^{k+1}}{(k+1)!},\quad
N_k^{(o)} := {\fkg k}\frac{t^k}{k!} \frac{{\bf n}_0\cdot{\bf r}_{ES}}{c},\quad
N_k^{(r)} := {\fkg k}\frac{t^k}{k!} \frac{{\bf n}_0\cdot{\bf r}_E}{c},
\ee
where $f_g$ and ${\fkg k}$ are respectively the frequency and the spindown 
parameters of the gravitational-wave signal. The moduli of the quantites $N_k$, 
$N_k^{(o)}$ and $N_k^{(r)}$ can be estimated as follows:
\bea
\label{ap16a}
&{\dst \left\vert N_k\right\vert = {\bar N}_k :=\bigg\vert{\fkg k}\bigg\vert
\frac{t^{k+1}}{(k+1)!},}& \\
\label{ap16b}
&{\dst  \left\vert N_k^{(o)}\right\vert \le{\bar N}_k^{(o)}
:= \bigg\vert{\fkg k}\bigg\vert \frac{t^k}{k!}\frac{r_{ES}}{c},\quad
\left\vert N_k^{(r)}\right\vert \le{\bar N}_k^{(r)}
:= \bigg\vert{\fkg k}\bigg\vert \frac{t^k}{k!}\frac{r_E}{c}. }&
\eea
We estimate the maximum values of the spindown parameters ${\fkg k}$ from the 
following formula
\be
\label{ap17}
\bigg\vert{\fkg k}\bigg\vert \simeq k!\frac{f_g}{\tau^k},
\ee
where $\tau$ is the spindown age of the neutron star. Using Eq.\ (\ref{ap17}) 
the quantities ${\bar N}_k$, ${\bar N}_k^{(o)}$ and ${\bar N}_k^{(r)}$ can be 
estimated as follows:
\bea
\label{ap18a}
&{\dst {\bar N}_k \simeq f_g\,t \frac{1}{k+1}\left(\frac{t}{\tau}\right)^k,}& \\
\label{ap18b}
&{\dst {\bar N}_k^{(o)} \simeq f_g\frac{r_{ES}}{c}\left(\frac{t}{\tau}\right)^k, 
\quad
{\bar N}_k^{(r)} \simeq f_g\frac{r_E}{c}\left(\frac{t}{\tau}\right)^k. }&
\eea
Let us note that the ratios of the different contributions for the fixed $k$ do 
not depend on the gravitational-wave frequency $f_g$ and the spindown age 
$\tau$:
\be
\label{ap19}
\frac{{\bar N}_k^{(r)}}{{\bar N}_k} \simeq (k+1)\frac{r_E}{c\,t},\quad
\frac{{\bar N}_k^{(o)}}{{\bar N}_k} \simeq (k+1)\frac{r_{ES}}{c\,t}.
\ee

From the formulae given above it is easy to obtain how many terms are needed in
each of the three series on the right-hand side of Eq.\ (\ref{ap200}) for the
eight models of the phase considered in our simulations in order to meet the
criterion that all terms that contribute more than 1/4 of a cycle are included.
The results are summarized in Table 1 of Sec.\ 2.

\section{Polynomial phase model}

The ability to control the values of the Fisher information matrix for the
frequency of the signal and its derivatives may be useful in real data
processing schemes. Therefore in this appendix we study the dependence of the 
Fisher matrix on the choice of the instant of time at which the instantaneous 
frequency and spindown parameters are defined.

We use the one-component signal with a constant amplitude and the phase 
polynomial 
in time, which is introduced in Sec.\ 5.4. Such a signal can be written as
\be
\label{poly1}
h(t;h_o,\boldsymbol{\zeta}) = h_o \sin\Psi(t;\boldsymbol{\zeta}),
\ee
where $h_o$ is a constant amplitude and $\boldsymbol{\zeta}$ 
denotes the parameters entering the phase $\Psi$ of the signal. 
As the signal (\ref{poly1}) has the constant amplitude, the formulae 
(\ref{ca2})--(\ref{ca6c}) from Sec.\ 6 can be applied here. But here unlike 
in Sec.\ 6 we choose the observation interval to be $[0,T_o]$, where $T_o$ is 
the observation time. We also introduce the dimensionless variable $x:=t/T_o$. 
Then the formulae (\ref{ca6a})--(\ref{ca6c}) for the components of the 
covariance matrix $C$ can be rewritten as
\bea
C_{h_oh_o} &\cong& \frac{h_o^2}{d^2},\\
C_{h_o\zeta_i} &\cong& 0,\\
C_{\zeta_i\zeta_j} &\cong& \frac{1}{d^2}
\left({\widetilde \Gamma}^{-1}\right)_{\zeta_i\zeta_j},
\eea
where the optimal signal-to-noise ratio $d$ is given by Eq.\ (\ref{ca2}) and the 
reduced Fisher matrix ${\widetilde \Gamma}$ is defined by
\be
\label{appb5}
{\widetilde \Gamma}_{\zeta_i\zeta_j} :=
\int_{0}^{1} \frac{\pa \Psi}{\pa \zeta_i}\frac{\pa \Psi}{\pa \zeta_j}\,dx.
\ee

Using the dimensionless variable $x$ the polynomial phase is defined as
\be
\label{appb6}
\Psi(x;\boldsymbol{\zeta}) = \Phi_0 + \sum_{k=0}^s \omega_k(x-x_0)^{k+1},
\ee
where $x_0$ is an arbitrarily chosen initial time, $\Phi_0$ denotes the initial 
phase, and $\omega_k$ ($k=0,\ldots,s$) are the spindown parameters, so 
$\boldsymbol{\zeta}=(\Phi_0,\omega_0,\ldots,\omega_s)$.

The reduced Fisher matrix ${\widetilde \Gamma}$ for the polynomial phase model 
(\ref{appb6}) has the structure very similar to that of the $(s+2)$-dimensional 
Hilbert matrix. For $x_0=0$ ${\widetilde \Gamma}$ is the 
$(s+2)$-dimensional Hilbert matrix. For $s=4$ (four spindown parameters 
included) the reduced Fisher matrix ${\widetilde \Gamma}$ computed for an 
arbitrary $x_0$ equals
\be
{\widetilde \Gamma} = \left(\begin{array}{cccccc}
\Gamma_1&\Gamma_2&\Gamma_3&\Gamma_4&\Gamma_5&\Gamma_6\\
&\Gamma_3&\Gamma_4&\Gamma_5&\Gamma_6&\Gamma_7\\
&&\Gamma_5&\Gamma_6&\Gamma_7&\Gamma_8\\
&&&\Gamma_7&\Gamma_8&\Gamma_9\\
&&&&\Gamma_9&\Gamma_{10}\\
&&&&&\Gamma_{11}
\end{array}\right),
\ee
where
\ben
\Gamma_1 &=& 1,\\
\Gamma_2 &=& 1/2 - x_0,\\
\Gamma_3 &=& 1/3 - x_0 + x_0^2,\\
\Gamma_4 &=& (1 - 4x_0 + 6x_0^2 - 4x_0^3)/4,\\
\Gamma_5 &=& 1/5 - x_0 + 2x_0^2 - 2x_0^3 + x_0^4,\\
\Gamma_6 &=& (1 - 6x_0 + 15x_0^2 - 20x_0^3 + 15x_0^4 - 6x_0^5)/6,\\
\Gamma_7 &=& 1/7 - x_0 + 3x_0^2 - 5x_0^3 + 5x_0^4 - 3x_0^5 + x_0^6,\\
\Gamma_8 &=& 1/8 - x_0 + (7x_0^2)/2 - 7x_0^3 + (35x_0^4)/4 - 7x_0^5 + 
(7x_0^6)/2 - x_0^7,\\
\Gamma_9 &=& 1/9 - x_0 + 4x_0^2 - (28x_0^3)/3 + 14x_0^4 - 14x_0^5 + 
(28x_0^6)/3 - 4x_0^7 + x_0^8,\\
\Gamma_{10} &=& 1/10 - x_0 + (9x_0^2)/2 - 12x_0^3 + 21x_0^4 - (126x_0^5)/5 + 
21x_0^6 - 12x_0^7 + (9x_0^8)/2 - x_0^9,\\
\Gamma_{11} &=& 1/11 - x_0 + 5x_0^2 - 15x_0^3 + 30x_0^4 - 42x_0^5 + 42x_0^6 - 
30x_0^7 + 15x_0^8 - 5x_0^9 + x_0^{10}.
\een
After putting $x_0=0$ into the above formulae one obtains the 6-dimensional 
Hilbert matrix:
\be
{\widetilde \Gamma} = \left(\begin{array}{cccccc}
1&1/2&1/3&1/4&1/5&1/6\\
1/2&1/3&1/4&1/5&1/6&1/7\\
1/3&1/4&1/5&1/6&1/7&1/8\\
1/4&1/5&1/6&1/7&1/8&1/9\\
1/5&1/6&1/7&1/8&1/9&1/10\\
1/6&1/7&1/8&1/9&1/10&1/11
\end{array}\right).
\ee

We have computed the rms errors of the phase parameters $\zeta_k$ for five
different phase models (\ref{appb6}) with $s=0,\ldots,4$.  The rms errors
$\sigma(\zeta_k)$ are the square roots of the diagonal elements of the reduced
covariance matrices ${\widetilde C}={\widetilde \Gamma}^{-1}$.  In Figure
\ref{f:cov} the errors $\sigma(\zeta_k)$ are plotted as functions of the initial
time $x_0$.  It is seen from the figure that the errors depend on the choice of
the initial time $x_0$.  This can be explained as follows.  The phase $\Psi$ of
the signal at a given time $x$ does not depend on the choice of the initial time
$x_0$.  Thus for the two different choices $x_{01}$ and $x_{02}$ of the initial
time $x_0$ we have
\be
\label{appb9}
\Phi_{01} + \sum_{k=0}^s \omega_{k1}(x-x_{01})^{k+1}
= \Phi_{02} + \sum_{k=0}^s \omega_{k2}(x-x_{02})^{k+1},
\ee
where the parameters $\Phi_{01},\omega_{k1}$ and $\Phi_{02},\omega_{k2}$ are 
associated with the initial times $x_{01}$ and $x_{02}$, respectively. Using 
Eq.\ (\ref{appb9}) one can, for fixed $s$, express the parameters 
$\boldsymbol{\zeta}_1=(\Phi_{01},\omega_{01},\ldots,\omega_{s1})$ as linear 
combinations of 
$\boldsymbol{\zeta}_2=(\Phi_{02},\omega_{02},\ldots,\omega_{s2})$:
\be
\label{appb10}
\zeta_{2l} = \sum_{n=0}^{s+2} j_{ln}(x_{01},x_{02}) \zeta_{1n},
\ee
where $j_{ln}$ are the components of the transformation matrix $J$ given by
\be
J = \left(\begin{array}{cccccc}
1 & x_{02}-x_{01} & (x_{02}-x_{01})^2 & (x_{02}-x_{01})^3 & (x_{02}-x_{01})^4
& (x_{02}-x_{01})^5 \\
0 & 1 & 2(x_{02}-x_{01}) & 3(x_{02}-x_{01})^2 & 4(x_{02}-x_{01})^3
& 5(x_{02}-x_{01})^4 \\ 
0 & 0 & 1 & 3(x_{02}-x_{01}) & 6(x_{02}-x_{01})^2 & 10(x_{02}-x_{01})^3 \\ 
0 & 0 & 0 & 1 & 4(x_{02}-x_{01}) & 10(x_{02}-x_{01})^2 \\
0 & 0 & 0 & 0 & 1 & 5(x_{02}-x_{01}) \\
0 & 0 & 0 & 0 & 0 & 1
\end{array}\right).
\ee
Thus the change of the initial time from $x_{01}$ to $x_{02}$ corresponds to
estimating a different set of parameters given by a linear combination of the
original parameters.  For example the new frequency parameter defined as the
first derivative of the phase at time $x_{02}$ (divided by $2\pi$) is a linear
combination of the frequency parameter and spindown parameters at time $x_{01}$.
Let us denote by $C(x_{01})$ and $C(x_{02})$ the covariance matrices for the
parameters $\zeta_{1l}$ and $\zeta_{2l}$, respectively.  It is easy to see that
these matrices are related by
\be
 C(x_{02}) = JC(x_{01})J^T.
\ee

The dependence of the rms errors of the signal parameters on the choice of the
initial time for the full phase model is very similar to that observed in the
polynomial model considered above.  This is illustrated in Figure \ref{f:hie},
where we have plotted the simulated cumulative distribution functions of the
parameter errors for the full model of the signal described in Sec.\ 2.  The
initial time $t_o=0$ used in the simulations coincides with the middle of the
observation interval which is $[-T_o/2,T_o/2]$, so it corresponds to $x_0=0.5$.
The comparison of Figures \ref{f:cov} and \ref{f:hie} shows that the curve
merging observed in Figure \ref{f:hie} exactly corresponds to intersection
points observed in Figure \ref{f:cov} for $x_0=0.5$.

Finally in Table \ref{T:pol} we give the rms errors (diagonal components of the
inverse of the Fisher matrix) of the parameters $\boldsymbol{\zeta}$ in various 
polynomial models.  We consider models with $s=0,\ldots,4$ spindowns and we
assume $x_0=0.5$.

\begin{table}\begin{center}
\begin{tabular}{|c|c|c|c|c|c|c|}\hline
$s$ & $\sigma({\Phi_0})d$ & $\sigma(\omega_0)d$ & $\sigma(\omega_1)d$ &
$\sigma(\omega_2)d$ & $\sigma(\omega_3)d$ & $\sigma(\omega_4)d$ \\
\hline\hline
0 & 1 & $2\sqrt{3}$ & & & & \\ \hline
1 & 3/2 & $2\sqrt{3}$ & $6\sqrt{5}$ & & & \\ \hline
2 & 3/2 & $5\sqrt{3}$ & $6\sqrt{5}$ & $20\sqrt{7}$ & & \\ \hline
3 & 15/8 & $5\sqrt{3}$ & $21\sqrt{5}$ & $20\sqrt{7}$ & $210$ & \\ \hline
4 & 15/8 & $35\sqrt{3}/4$ & $21\sqrt{5}$ & $90\sqrt{7}$ & $210$ & $252\sqrt{11}$ 
\\ \hline
\end{tabular}\end{center}
\caption{\label{T:pol}
The rms errors $\sigma(\zeta_k)$ of the phase parameters 
$\boldsymbol{\zeta}=(\Phi_0,\omega_0,\ldots,\omega_4)$ for the polynomial models 
with $s=0,\ldots,4$ spindowns. We have assumed the initial time $x_0=0.5$.}
\end{table}


\unitlength 1cm
\begin{center}\begin{figure}[ht]
\begin{picture}(10,20)
\put(0,0){\includegraphics{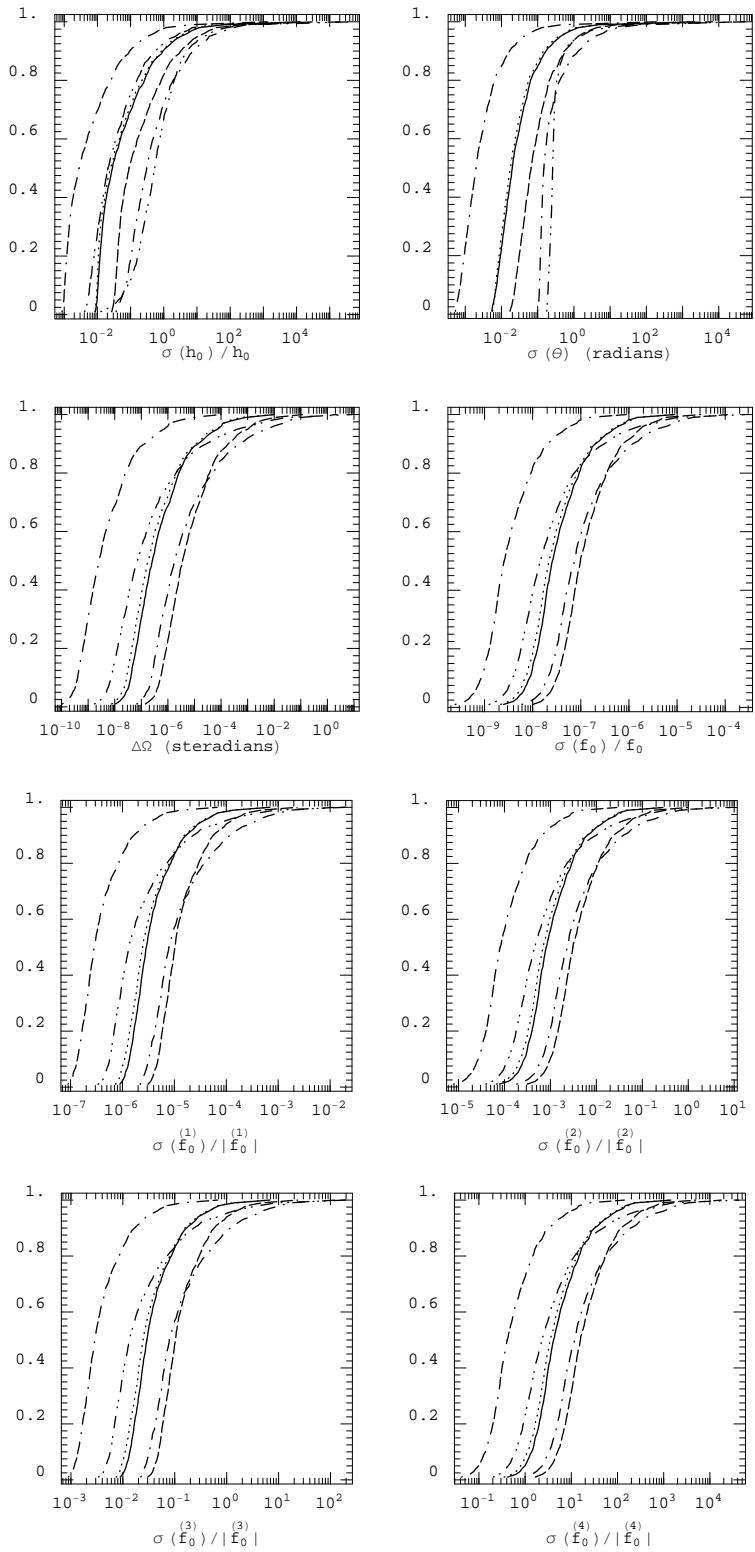}}
\end{picture}
\caption{ Cumulative distribution functions of the simulated rms errors of the
signal parameters for the individual detectors in the case of all-sky searches.
The observation time $T_o=120$ days.  We assume that star's ellipticity
$\epsilon=10^{-5}$, its moment of inertia w.r.t.\ the rotation axis $I=10^{45}$
g cm$^2$, its distance from the Earth $r_o=1$ kpc, and the frequency
$f_o=500$ Hz [these values give the dimensionless amplitude of the waveform
$h_o=1.1\times10^{-23}$, cf.\ Eq.\ (\ref{ho})].  The spindown age $\tau=40$
years.  The model of the signal's phase is described by $s_1=4$, $s_2=3$, and
$s_3=0$ (cf.\ Table 1).  The lines given on the diagrams correspond to various
detectors:  advanced Hanford (dotted/double dashed), initial Hanford (solid),
VIRGO (dotted), wideband GEO600 (dashed), narrowband GEO600 (double
dotted/dashed), and TAMA300 (dotted/dashed).  On the left top panel we have also
put the extra double dotted/double dashed curve.  It describes cumulative
distribution function of the relative rms error of the amplitude
$h'_o=h_o\sin^2\theta$ of the second component of the signal as measured by the
GEO600 narrowband detector.}  \end{figure}\end{center}

\begin{center}\begin{figure}[ht]
\begin{picture}(10,21)
\put(0,0){\includegraphics{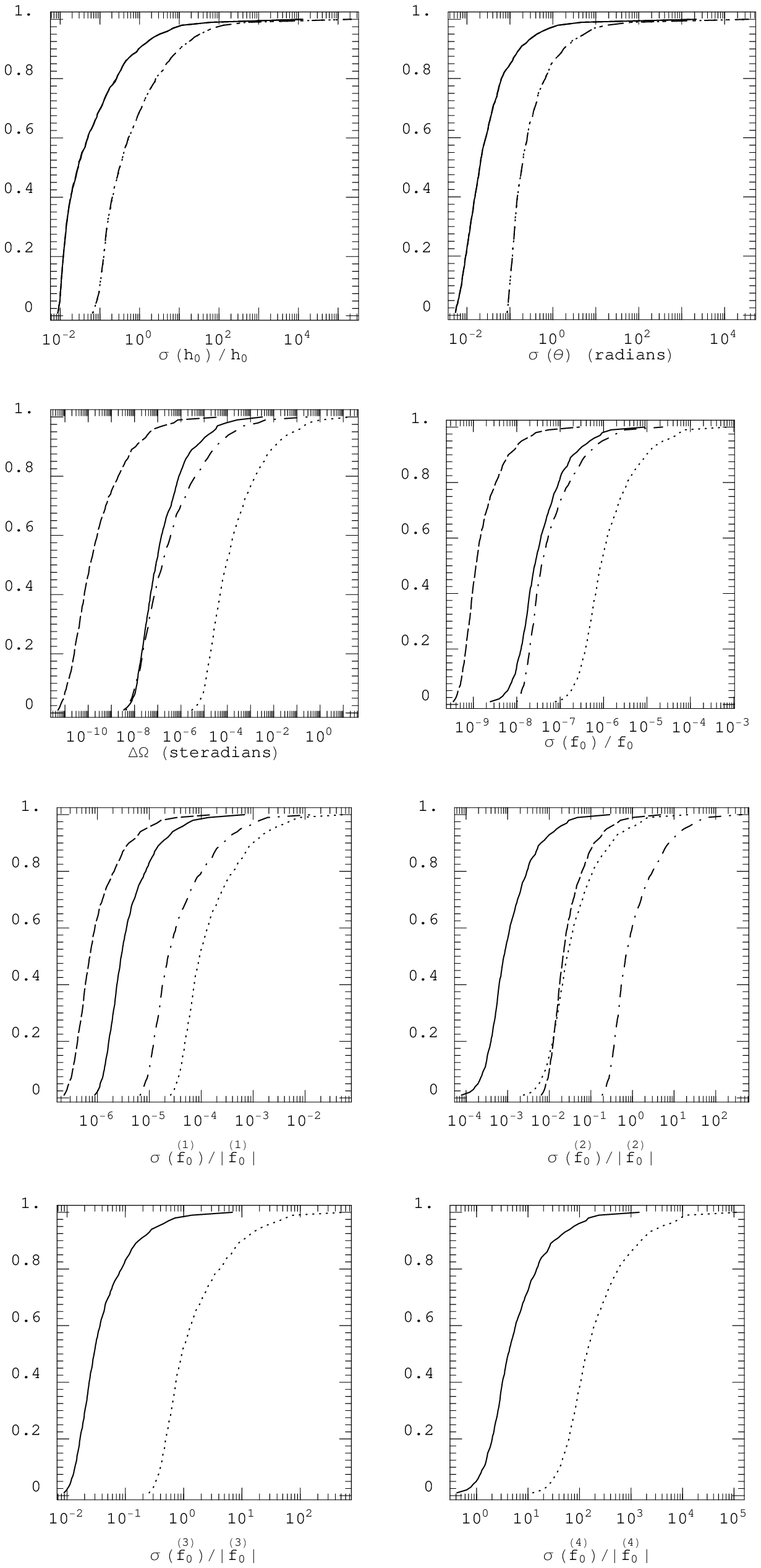}}
\end{picture}
\caption{Cumulative distribution functions of the simulated rms errors of the 
signal parameters for various models of the signal in the case of all-sky 
searches. The observation time $T_o=120$ days. The initial Hanford detector is 
assumed in the simulations. The neutron star parameters are the same as in 
Figure 1. The lines shown in the diagrams correspond to the following models of 
the signal:
$f_o=500$ Hz, $\tau=40$ yr,   $s_1=4$, $s_2=3$, $s_3=0$ (solid),
$f_o=500$ Hz, $\tau=1000$ yr, $s_1=2$, $s_2=1$, $s_3=0$ (dashed),
$f_o=100$ Hz, $\tau=40$ yr,   $s_1=4$, $s_2=2$, $s_3=0$ (dotted),
$f_o=100$ Hz, $\tau=1000$ yr, $s_1=2$, $s_2=1$, $s_3=0$ (dotted/dashed).
The dimensionless amplitude of the waveform $h_o=1.1\times10^{-23}$ for the 
models with $f_o=500$ Hz and $h_o=4.2\times10^{-25}$ for the models with 
$f_o=100$ Hz. On the two top panels the lines which correspond to the models 
with the same frequency $f_o$ coincide.}
\end{figure}\end{center}

\begin{center}\begin{figure}[ht]
\begin{picture}(10,20)
\put(0,0){\includegraphics{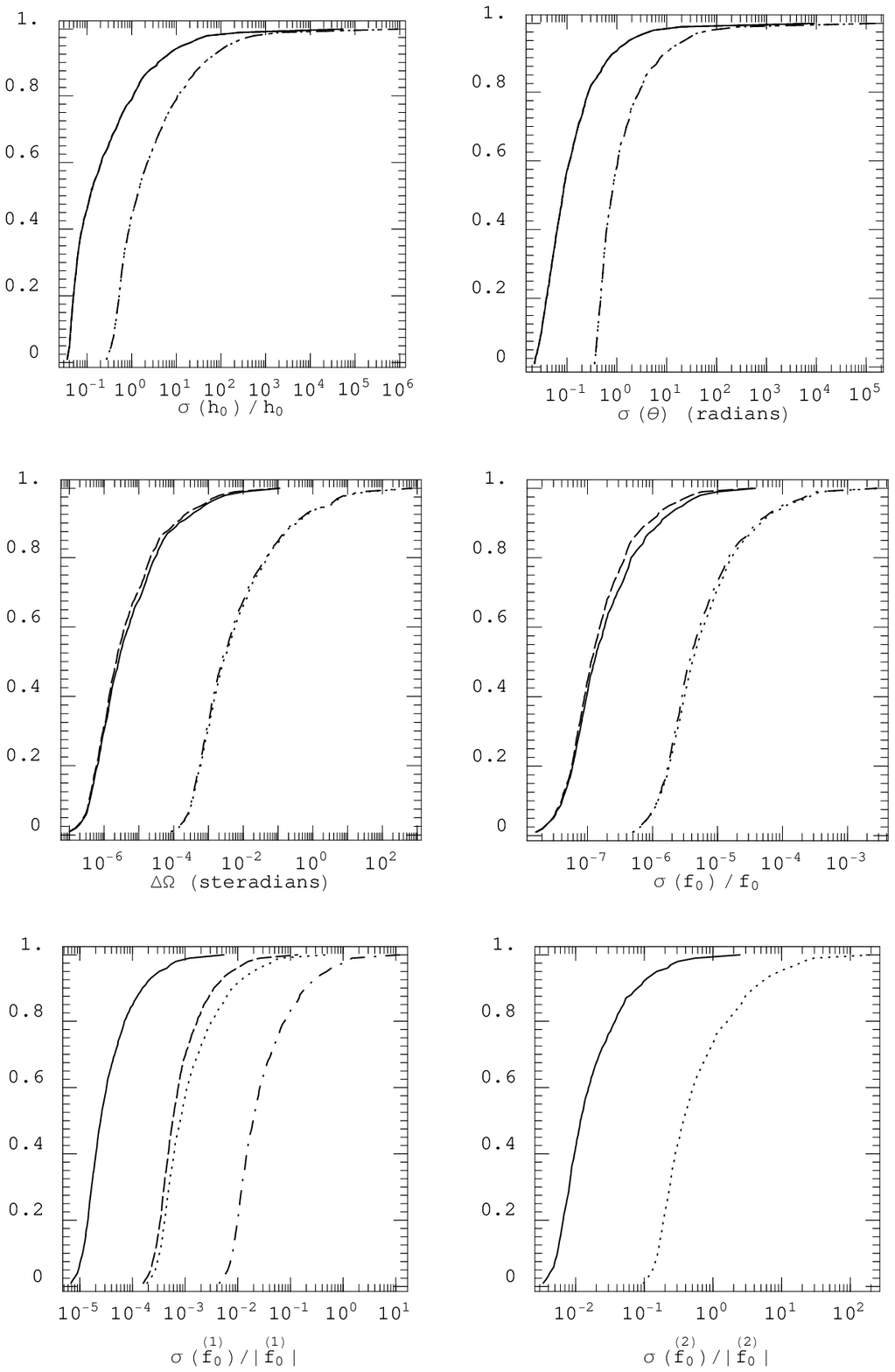}}
\end{picture}
\caption{Cumulative distribution functions of the simulated rms errors of the 
signal parameters for various models of the signal in the case of all-sky 
searches. The observation time $T_o=7$ days. The initial Hanford detector is 
assumed in the simulations. The neutron star parameters are the same as in 
Figure 1. The lines shown in the diagrams correspond to the following models of 
the signal:
$f_o=500$ Hz, $\tau=40$ yr,   $s_1=2$, $s_2=1$, $s_3=0$ (solid),
$f_o=500$ Hz, $\tau=1000$ yr, $s_1=1$, $s_2=1$, $s_3=0$ (dashed),
$f_o=100$ Hz, $\tau=40$ yr,   $s_1=2$, $s_2=1$, $s_3=0$ (dotted),
$f_o=100$ Hz, $\tau=1000$ yr, $s_1=1$, $s_2=1$, $s_3=0$ (dotted/dashed).
The dimensionless amplitude of the waveform $h_o=1.1\times10^{-23}$ for the 
models with $f_o=500$ Hz and $h_o=4.2\times10^{-25}$ for the models with 
$f_o=100$ Hz. On the two top panels the lines which correspond to the models 
with the same frequency $f_o$ coincide.}
\end{figure}\end{center}

\begin{center}\begin{figure}[ht]
\begin{picture}(10,21)
\put(0,0){\includegraphics{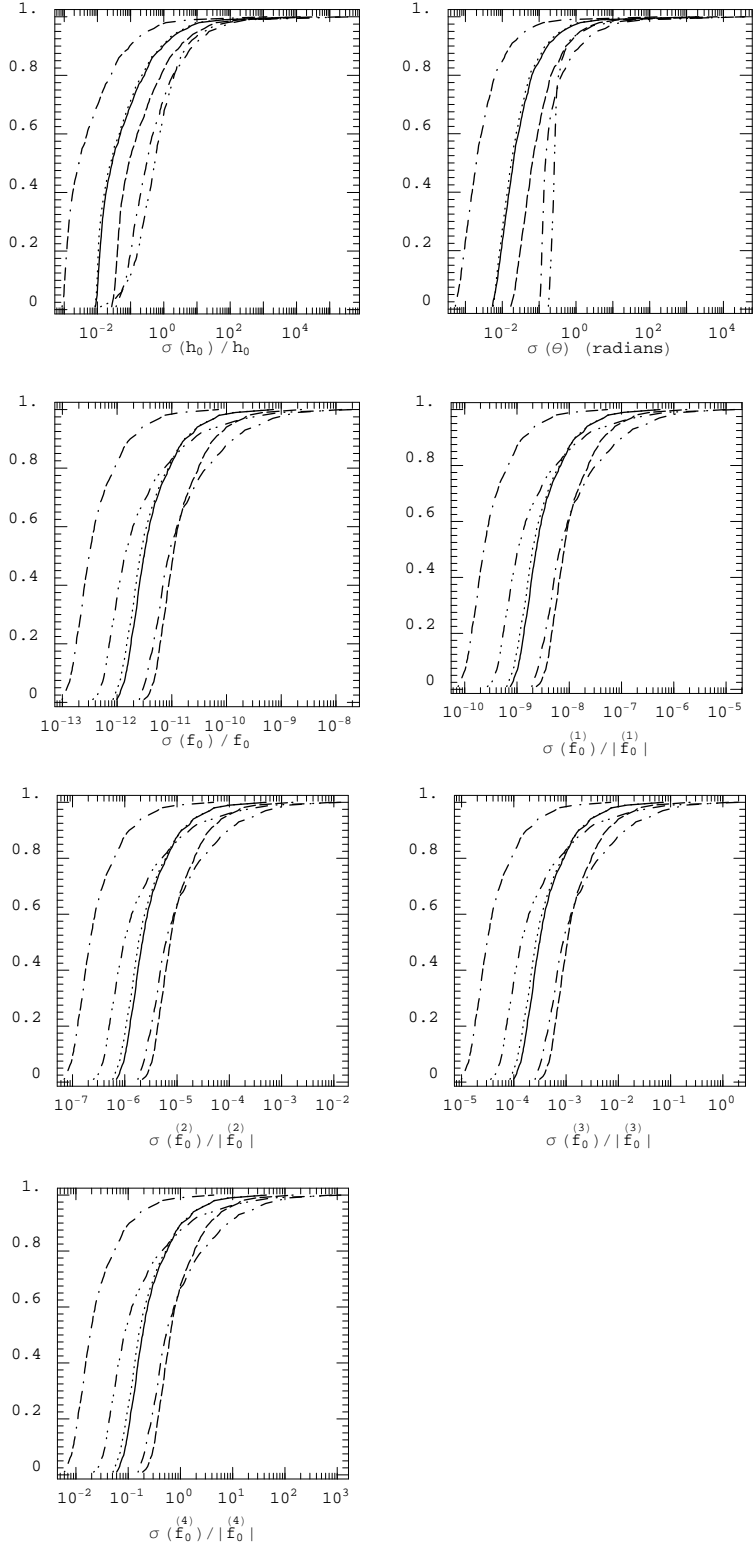}}
\end{picture}
\caption{Cumulative distribution functions of the simulated rms errors of the
signal parameters for the individual detectors in the case of directed searches.
The observation time $T_o=120$ days.  The neutron star parameters are the same
as in Figure 1.  The model of the signal's phase is described by $s_1=4$,
$s_2=3$, and $s_3=0$.  The lines given on the diagrams correspond to various
detectors:  advanced Hanford (dotted/double dashed), initial Hanford (solid),
VIRGO (dotted), wideband GEO600 (dashed), narrowband GEO600 (double
dotted/dashed), and TAMA300 (dotted/dashed).}  \end{figure}\end{center}

\begin{center}\begin{figure}[ht]
\begin{picture}(10,21)
\put(0,0){\includegraphics{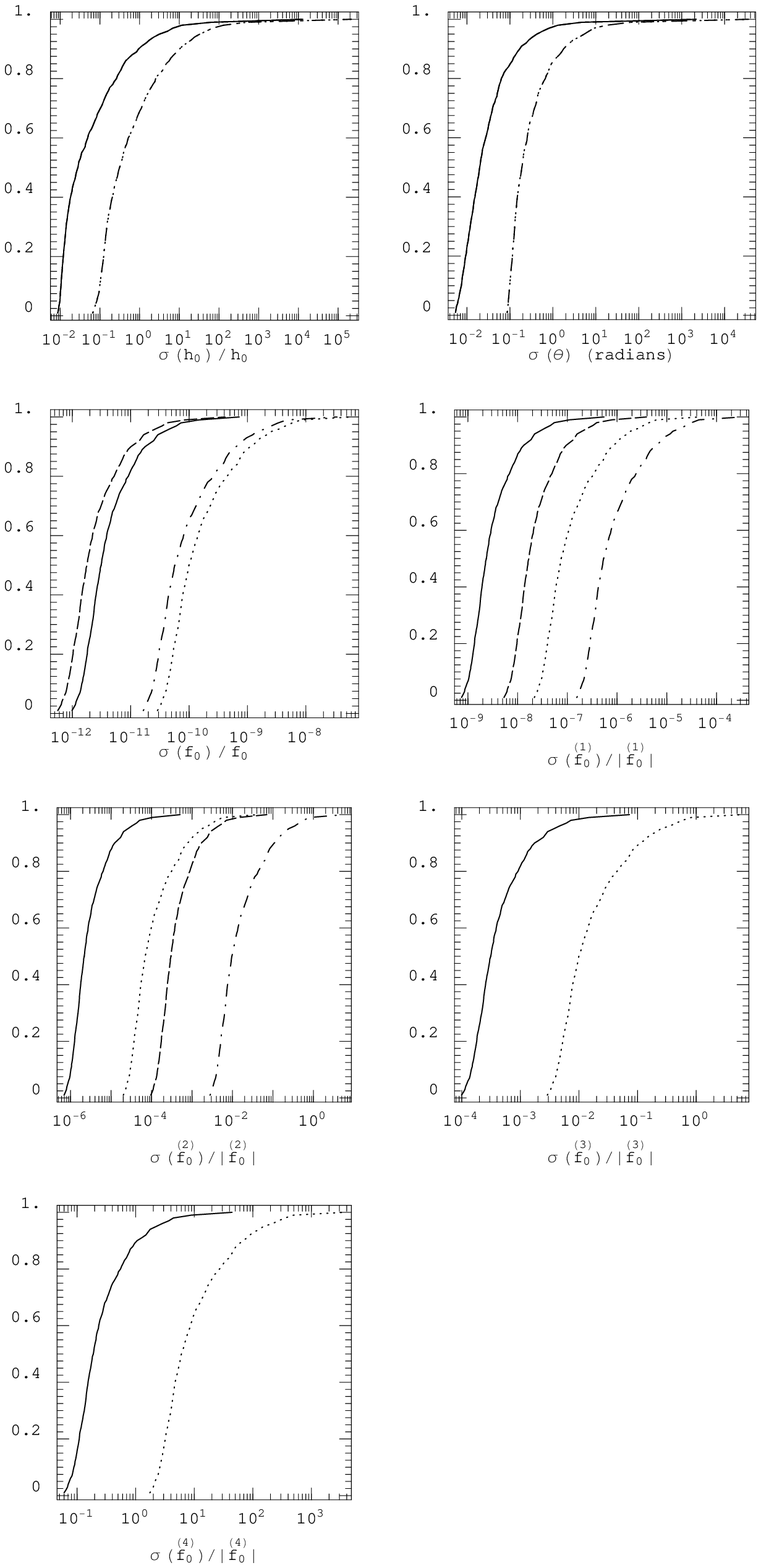}}
\end{picture}
\caption{Cumulative distribution functions of the simulated rms errors of the 
signal parameters for directed searches and various models of the signal.
The observation time $T_o=120$ days. The initial Hanford 
detector is assumed in the simulations. The neutron star parameters are the same 
as in Figure 1. The lines shown in the diagrams correspond to the following 
models of the parameter space:
$f_o=500$ Hz, $\tau=40$ yr,   $s_1=4$, $s_2=3$, $s_3=0$ (solid),
$f_o=500$ Hz, $\tau=1000$ yr, $s_1=2$, $s_2=1$, $s_3=0$ (dashed),
$f_o=100$ Hz, $\tau=40$ yr,   $s_1=4$, $s_2=2$, $s_3=0$ (dotted),
$f_o=100$ Hz, $\tau=1000$ yr, $s_1=2$, $s_2=1$, $s_3=0$ (dotted/dashed).
The dimensionless amplitude of the waveform $h_o=1.1\times10^{-23}$ for the 
models with $f_o=500$ Hz and $h_o=4.2\times10^{-25}$ for the models with 
$f_o=100$ Hz. On the two top panels the lines which correspond to the models 
with the same frequency $f_o$ coincide.}
\end{figure}\end{center}

\begin{center}\begin{figure}[ht]
\begin{picture}(10,20)
\put(0,0){\includegraphics{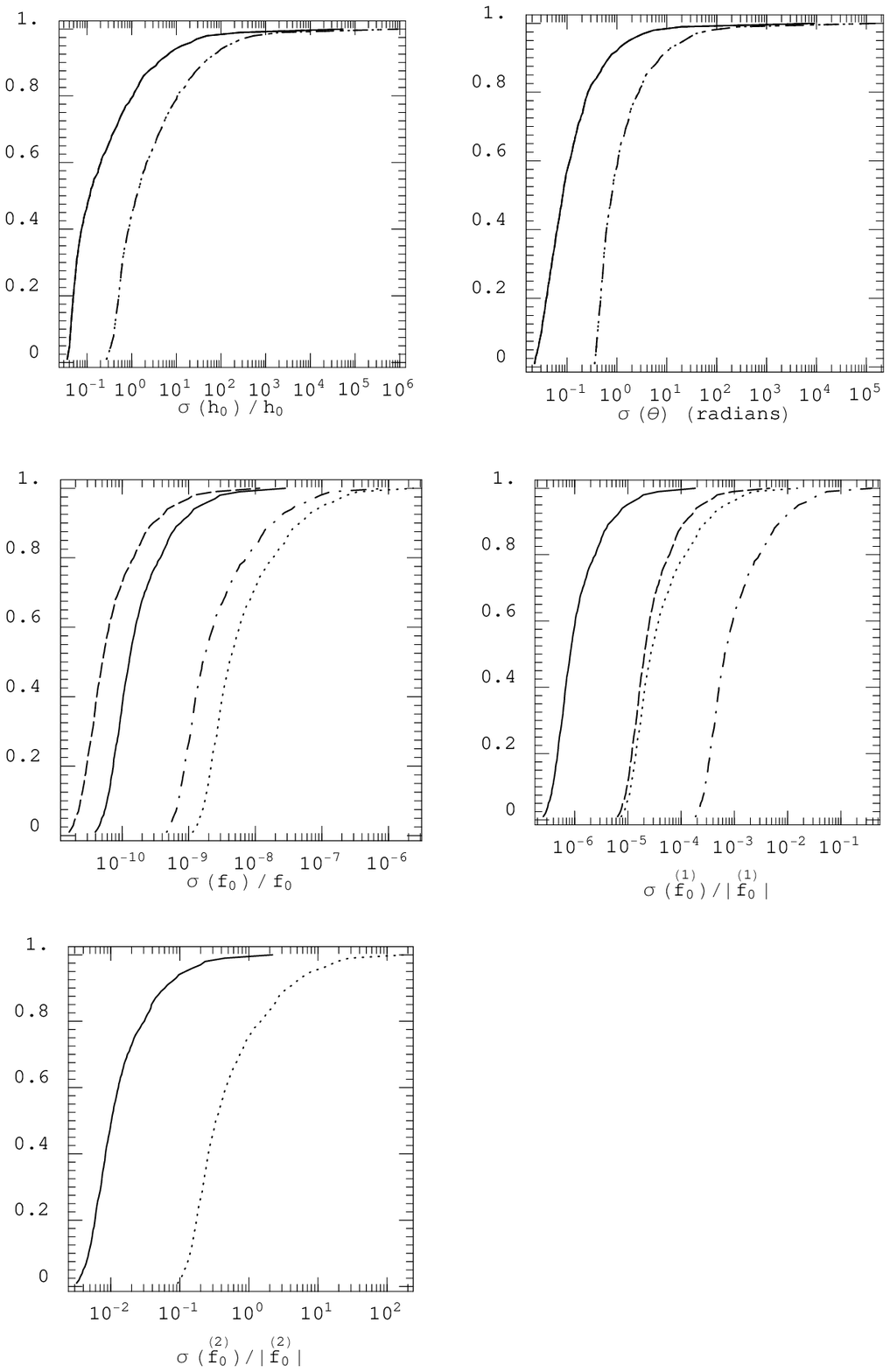}}
\end{picture}
\caption{Cumulative distribution functions of the simulated rms errors of the 
signal parameters for directed searches and various models of the signal.
The observation time $T_o=7$ days.  The initial Hanford 
detector is assumed in the simulations. The neutron star parameters are the same 
as in Figure 1. The lines shown in the diagrams correspond to the following 
models of the parameter space:
$f_o=500$ Hz, $\tau=40$ yr,   $s_1=2$, $s_2=1$, $s_3=0$ (solid),
$f_o=500$ Hz, $\tau=1000$ yr, $s_1=1$, $s_2=1$, $s_3=0$ (dashed),
$f_o=100$ Hz, $\tau=40$ yr,   $s_1=2$, $s_2=1$, $s_3=0$ (dotted),
$f_o=100$ Hz, $\tau=1000$ yr, $s_1=1$, $s_2=1$, $s_3=0$ (dotted/dashed).
The dimensionless amplitude of the waveform $h_o=1.1\times10^{-23}$ for the 
models with $f_o=500$ Hz and $h_o=4.2\times10^{-25}$ for the models with 
$f_o=100$ Hz. On the two top panels the lines which correspond to the models 
with the same frequency $f_o$ coincide.}
\end{figure}\end{center}

\begin{center}\begin{figure}[ht]
\begin{picture}(10,20)
\put(0,0){\includegraphics{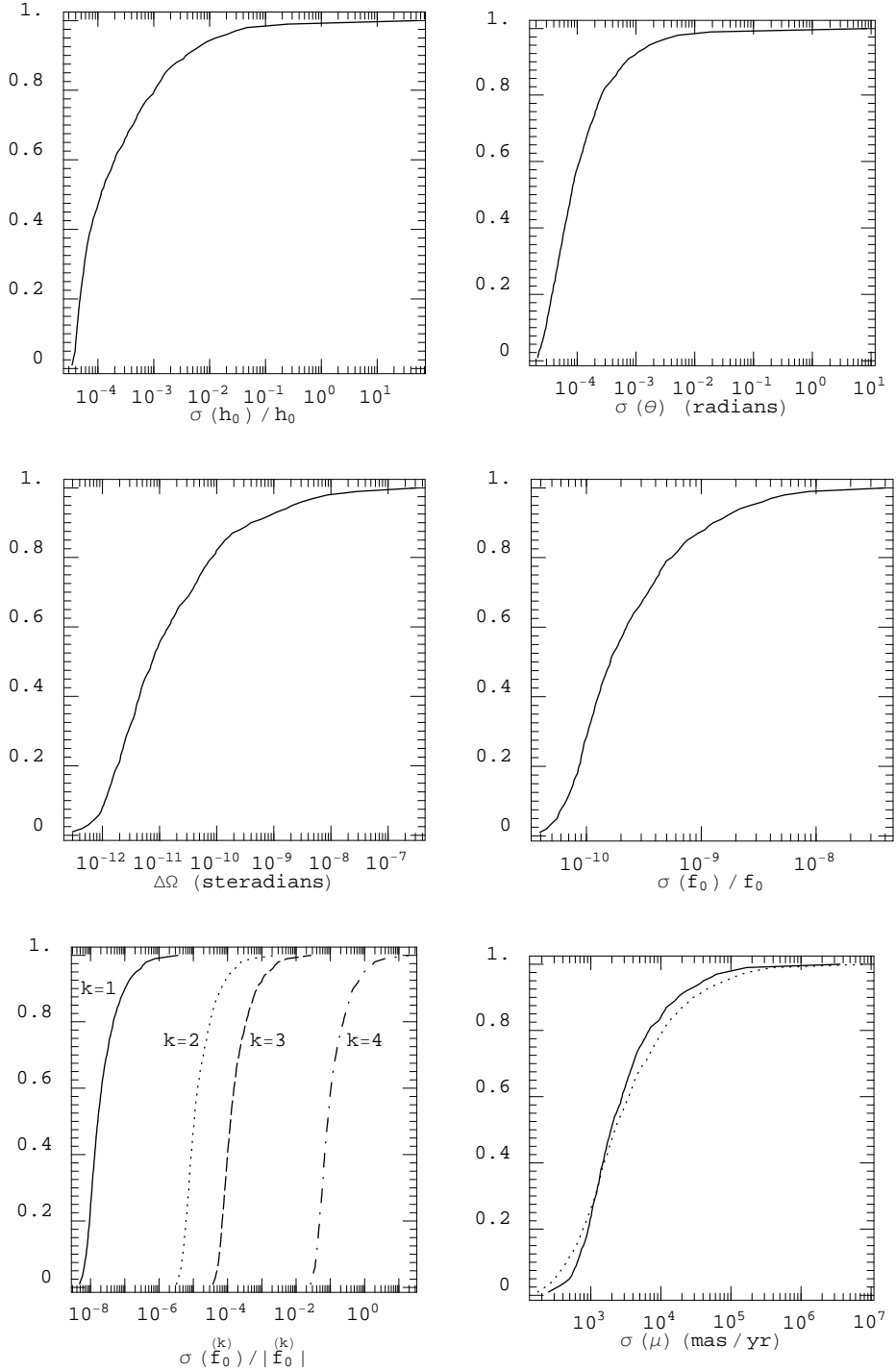}}
\end{picture}
\caption{Cumulative distribution functions of the simulated rms errors of the
signal parameters for all-sky searches including the two proper motion
parameters $\mu_{\alpha}$ and $\mu_{\delta}$.  The observation time $T_o=120$
days. The advanced Hanford detector was taken in the simulation.  We assume
that star's ellipticity $\epsilon=10^{-5}$, its moment of inertia w.r.t.\ the
rotation axis $I=10^{45}$ g cm$^2$, its distance from the Earth $r_o=40$ pc, and
the frequency $f_o=500$ Hz [these values give the dimensionless amplitude of the
waveform $h_o=2.6\times10^{-22}$].  The spindown age $\tau=40$ years.  We have 
assumed that neutron star transverse velocity is $10^3$ km/s. In the
bottom right panel we give the distributions of the rms errors 
$\sigma(\mu_{\alpha})$ (solid) and $\sigma(\mu_{\delta})$ (dotted).}  
\end{figure}\end{center}

\begin{center}\begin{figure}[ht]
\begin{picture}(10,9)
\put(0,0){\includegraphics{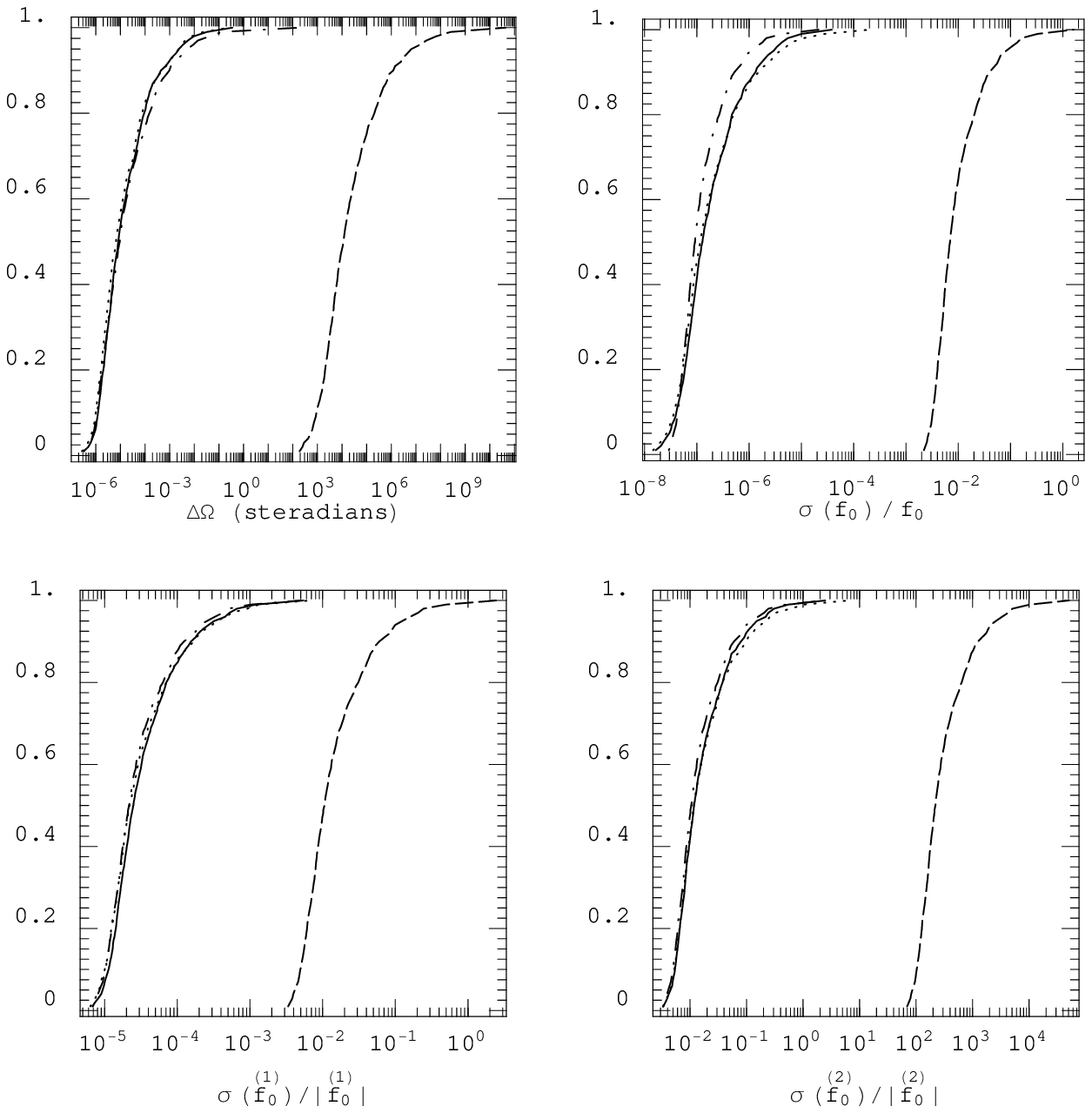}}
\end{picture}
\caption{\label{f:comp7}
Comparison of cumulative distribution functions of the simulated rms
errors of the signal parameters in the case of all-sky searches for the three
simplified models of the gravitational-wave signal with the signal model given
in Section 2.  The observation time $T_o=7$ days.  The
initial Hanford detector is assumed in the simulations.  The neutron star
parameters are the same as in Figure 1 so the solid lines are the same as solid
lines in Figure 1.  The other lines are as follows:  the constant amplitude
model (dotted), the linear model I (dotted/dashed), and the linear model II
(dashed).}  \end{figure}\end{center}

\begin{center}\begin{figure}[ht]
\begin{picture}(10,15)
\put(0,0){\includegraphics{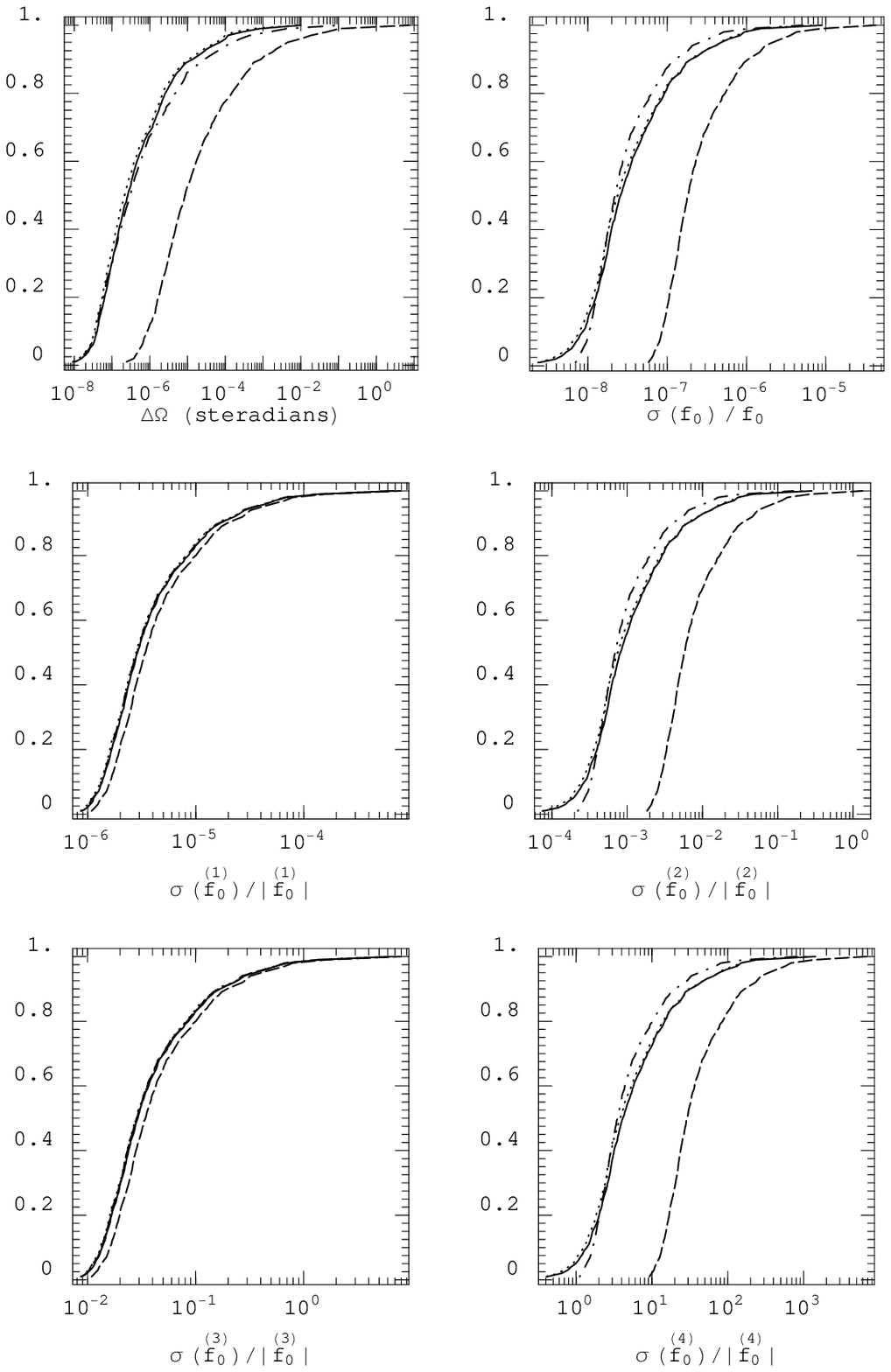}}
\end{picture}
\caption{\label{f:comp120}
Comparison of cumulative distribution functions of the simulated rms
errors of the signal parameters in the case of all-sky searches for the three
simplified models of the gravitational-wave signal with signal model given
in Section 2.  The observation time $T_o=120$ days.  The
initial Hanford detector is assumed in the simulations.  The neutron star
parameters are the same as in Figure 1 so the solid lines are the same as solid
lines in Figure 1.  The other lines are as follows:  the constant amplitude
model (dotted), the linear model I (dotted/dashed), and the linear model II
(dashed).}
\end{figure}\end{center}

\begin{center}
\begin{figure}[ht]
\begin{picture}(10,21)
\put(0,0){\includegraphics{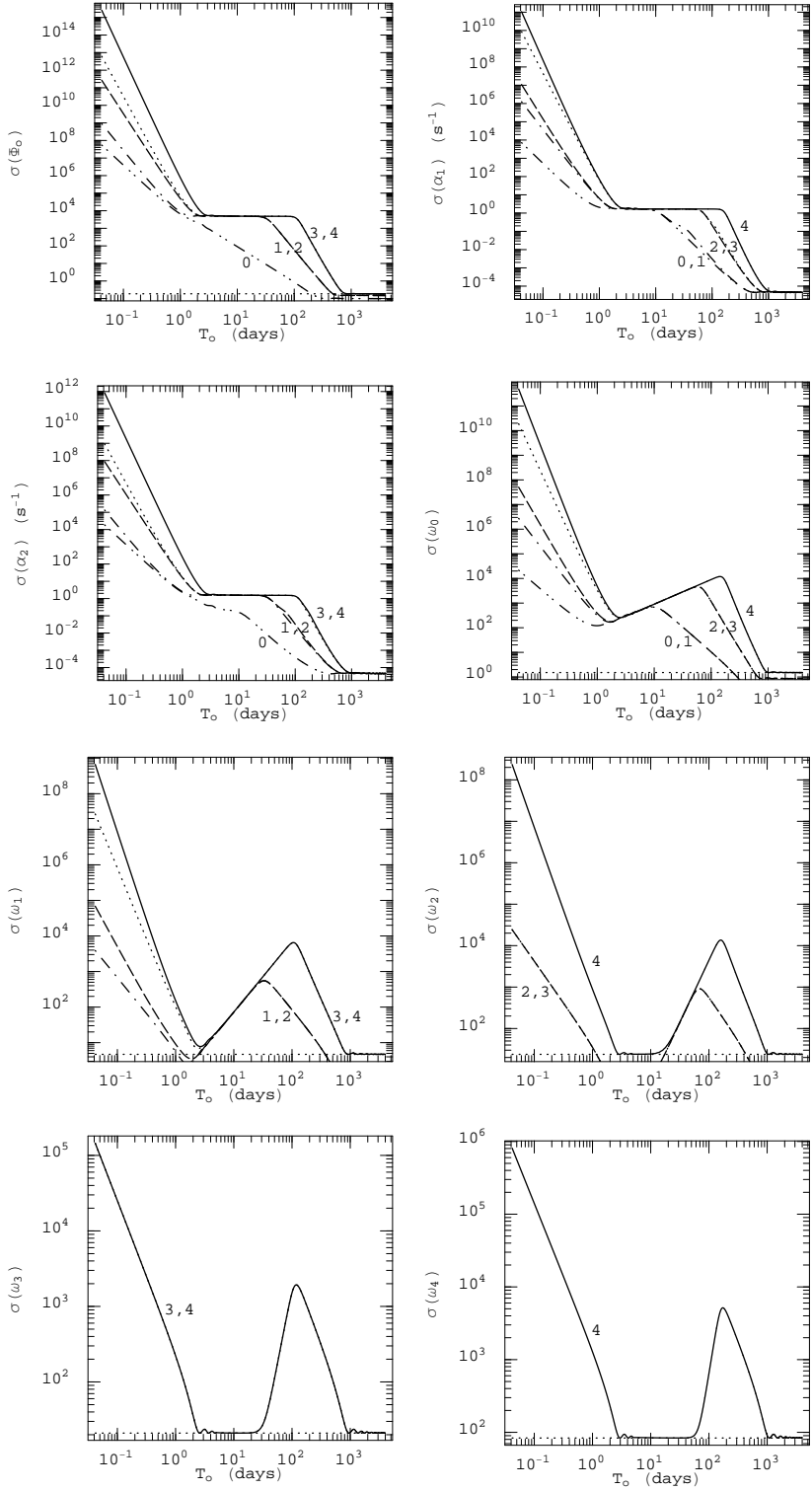}}
\end{picture}
\caption{\label{f:liot}
Dependence of the rms errors of the phase parameters on the observation time 
$T_o$. We have used the constant amplitude linear phase signal defined in Eqs.\ 
(\ref{sig6a}) and (\ref{sig6b}). The signal-to-noise ratio $d=10$. We have 
assumed the latitude $\lambda$ of the Hanford LIGO detector, we have also put 
$\phi_r=1.456$ and $\phi_o=0.123$. The lines (and the numbers) in the diagrams 
correspond to different numbers $s$ of spindown parameters included in the phase 
(\ref{sig6b}): $s=4$ (solid), $s=3$ (dotted), $s=2$ (dashed), $s=1$ 
(dotted/dashed), and $s=0$ (double dotted/dashed). The horizontal dotted lines 
give the rms errors of the parameters as predicted by the polynomial phase model 
with $s=4$ spindowns included (see Appendix C). The numbers corresponding to 
these lines are taken from Table 4 of Appendix C (for the signal-to-noise ratio 
$d=10$). We see that for the observation times $T_o\gtrsim1000$ days the 
horizontal lines coincide with the solid lines.} 
\end{figure}
\end{center}

\begin{center}\begin{figure}[ht]
\begin{picture}(10,21)
\put(0,0){\includegraphics{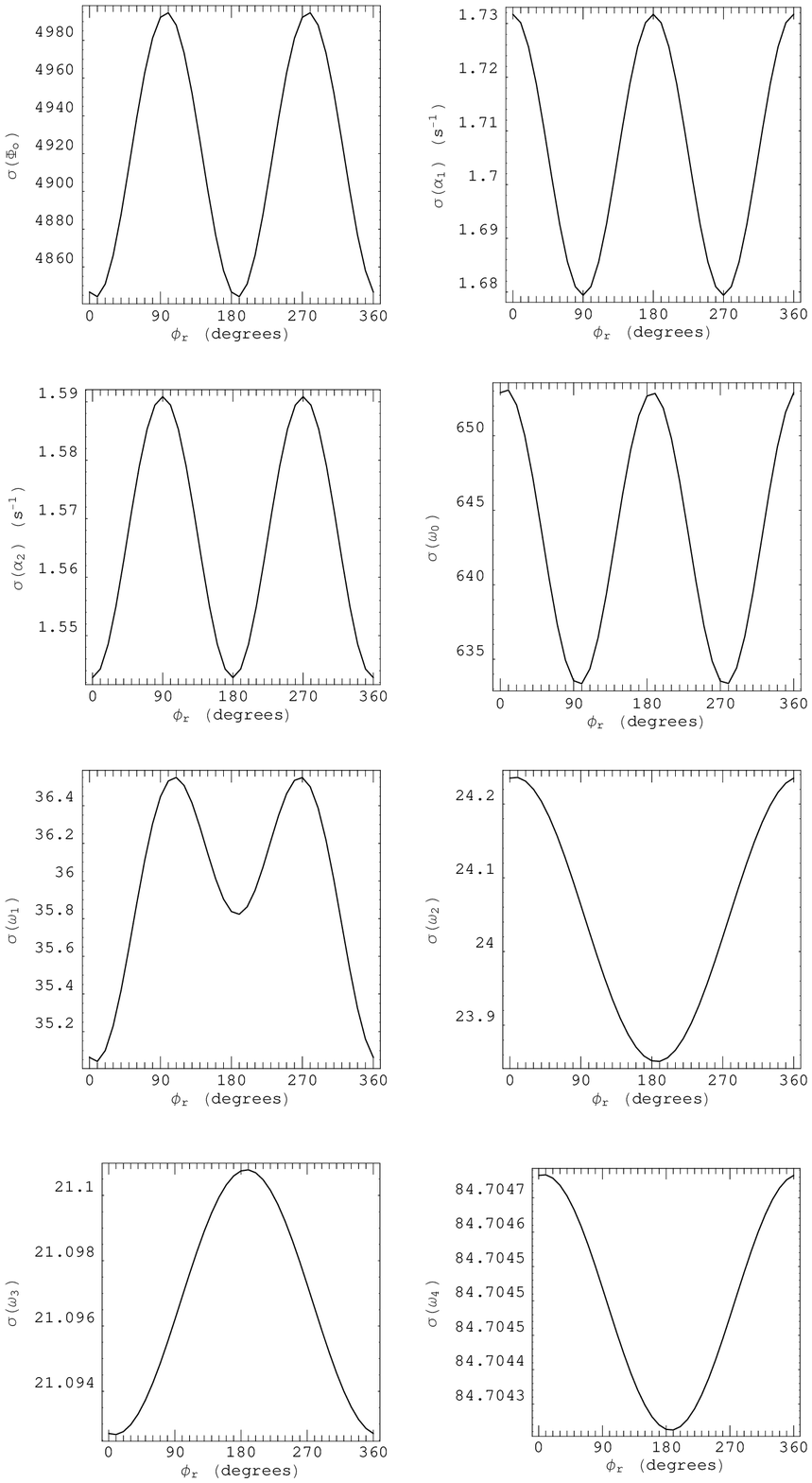}}
\end{picture}
\caption{\label{f:lirot}
Dependence of the rms errors of the phase parameters on the initial phase 
$\phi_r$ of the Earth's diurnal motion. We have used the constant amplitude 
linear phase signal defined in Eqs.\ (\ref{sig6a}) and (\ref{sig6b}) with $s=4$ 
spindown parameters included. The observation time $T_o$ = 7 days and the 
signal-to-noise ratio $d=10$. We have assumed the latitude $\lambda$ of the 
Hanford LIGO detector, we have also put $\phi_o=0.123$.}
\end{figure}\end{center}

\begin{center}\begin{figure}[ht]
\begin{picture}(10,21)
\put(0,0){\includegraphics{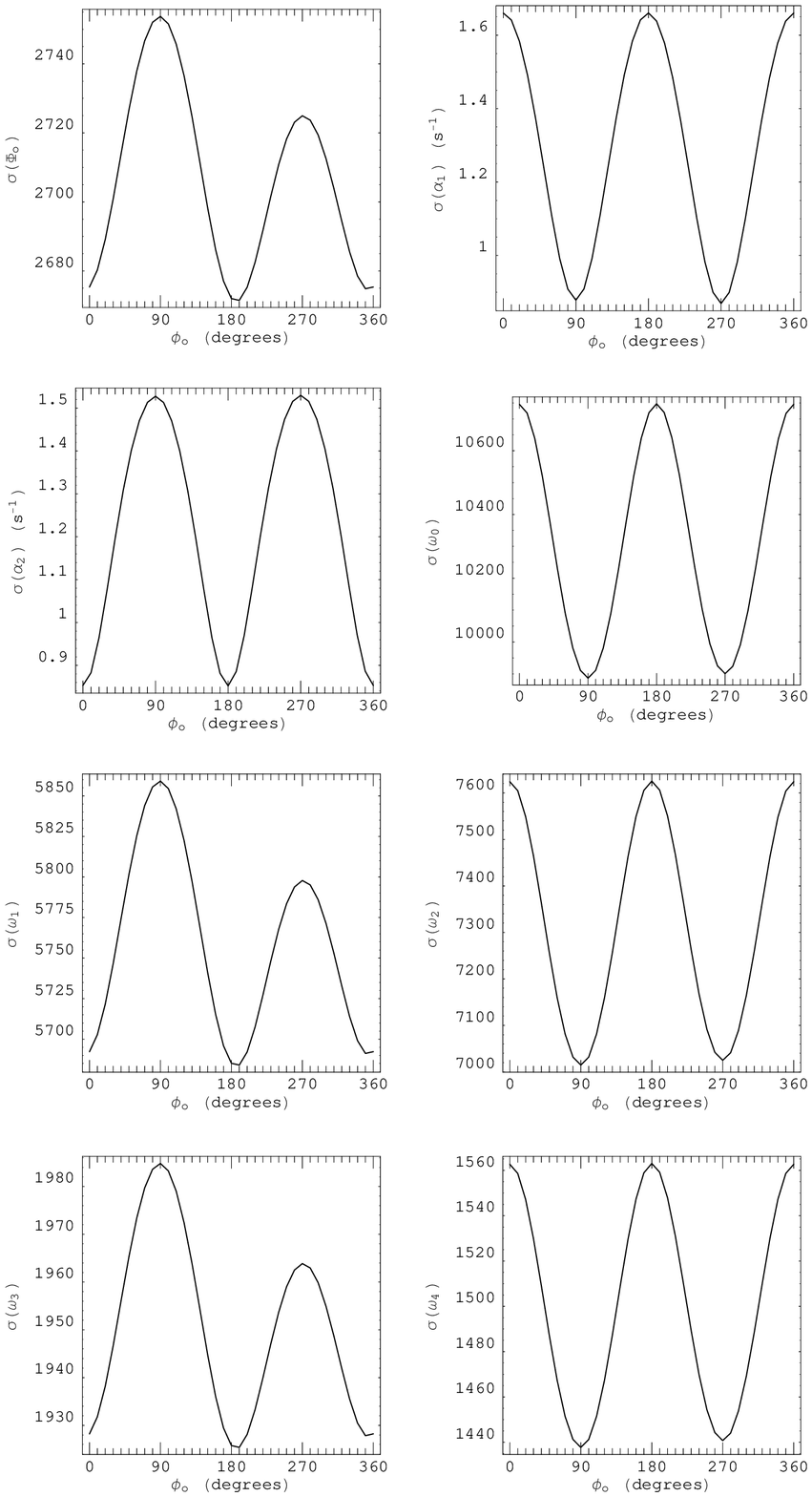}}
\end{picture}
\caption{\label{f:liorb}
Dependence of the rms errors of the phase parameters on the initial phase 
$\phi_o$ of the Earth's orbital motion. We have used the constant amplitude 
linear phase signal defined in Eqs.\ (\ref{sig6a}) and (\ref{sig6b}) with $s=4$ 
spindown parameters included. The observation time $T_o$ = 120 days and the 
signal-to-noise ratio $d=10$. We have assumed the latitude $\lambda$ of the 
Hanford LIGO detector, we have also put $\phi_r=1.456$.}
\end{figure}\end{center}

\begin{center}\begin{figure}[ht]
\begin{picture}(10,21)
\put(0,0){\includegraphics{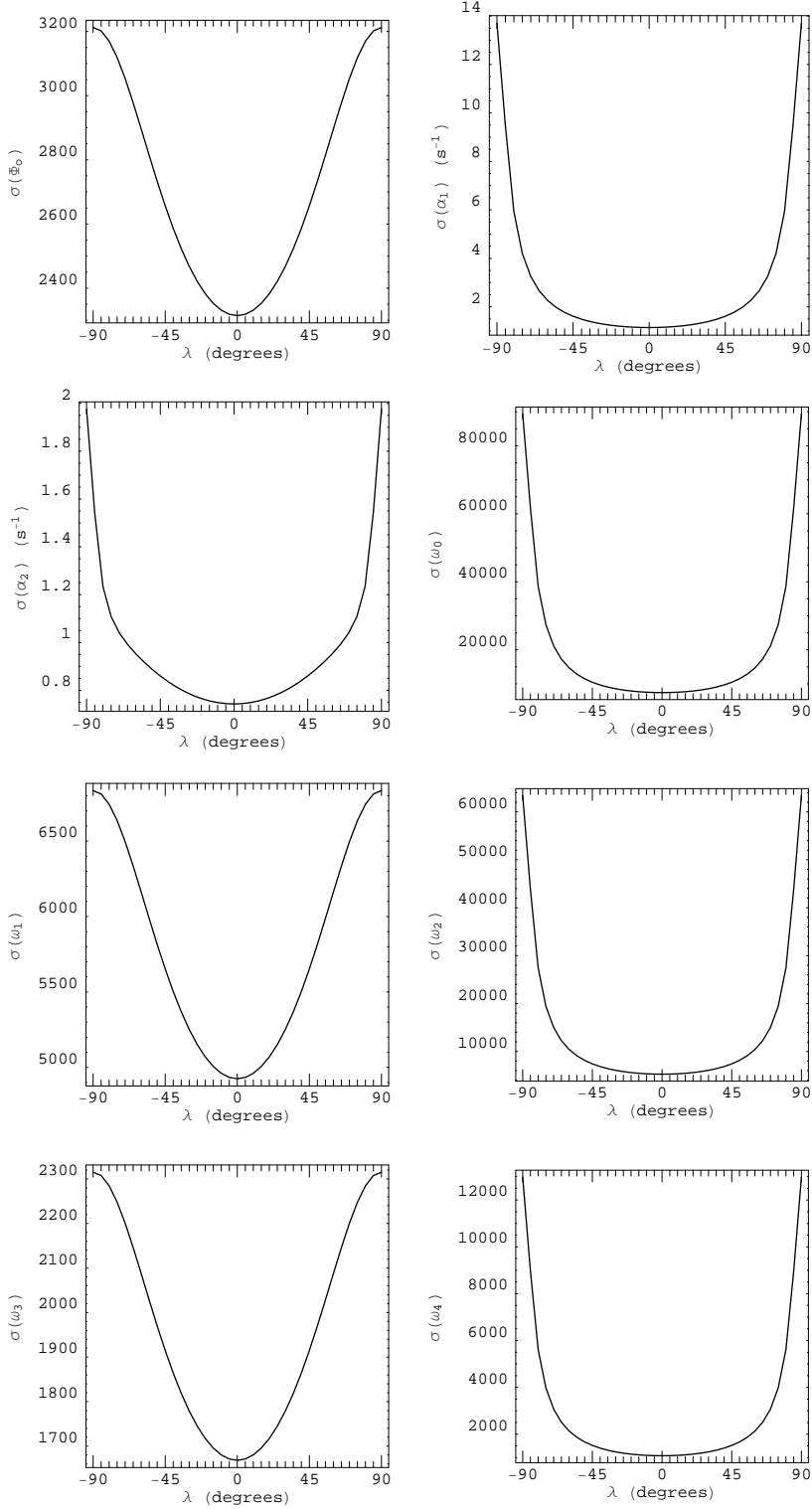}}
\end{picture}
\caption{\label{f:lilat}
Dependence of the rms errors of the phase parameters on the latitude $\lambda$ 
of the detector's location. We have used the constant amplitude linear phase 
signal defined in Eqs.\ (\ref{sig6a}) and (\ref{sig6b}) with $s=4$ spindown 
parameters included. The observation time $T_o$ = 120 days and the 
signal-to-noise ratio $d=10$. We have put $\phi_r=1.456$ and $\phi_o=0.123$.}
\end{figure}\end{center}

\begin{center}\begin{figure}[ht]
\begin{picture}(10,18)
\put(0,0){\includegraphics{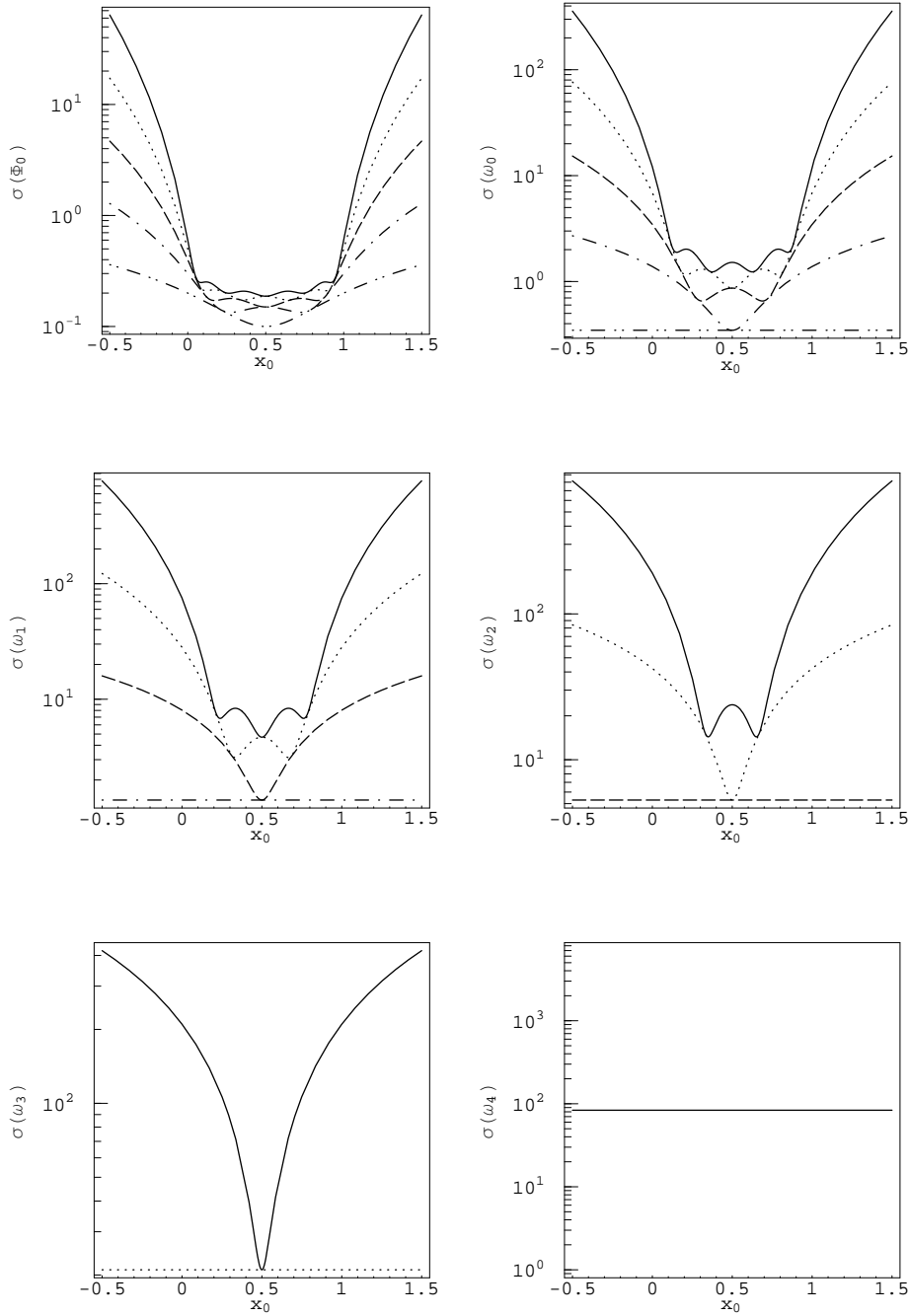}}
\end{picture}
\caption{\label{f:cov}
The rms errors $\sigma(\zeta_k)$ of the phase parameters 
$\boldsymbol{\zeta}=(\Phi_0,\omega_0,\ldots,\omega_4)$ plotted as functions of 
the initial time $x_0$ for the signal-to-noise ratio $d=10$. 
The curves correspond to models of the phase with 
increasing number of spindown parameters: dashed/double dotted curve---no 
spindown paremeter (this corresponds to a monochromatic signal with unknown 
constant initial phase), dashed/dotted curve---1 spindown parameter, dashed 
curve---2 spindown parameters, dotted---3 spindown parameters, solid---4 
spindown parameters. All the curves shown here are symmetric w.r.t.\ $x_0=0.5$. 
For some discrete values of $x_0$ curves for $(r+1)$-parameter model intersect 
with $r$-parameter model curves. This means that in such cases adding $r + 1$ 
parameter does not change the rms error in $r$th parameter.} 
\end{figure}\end{center}

\begin{center}\begin{figure}[ht]
\begin{picture}(10,20.9)
\put(0,0){\includegraphics{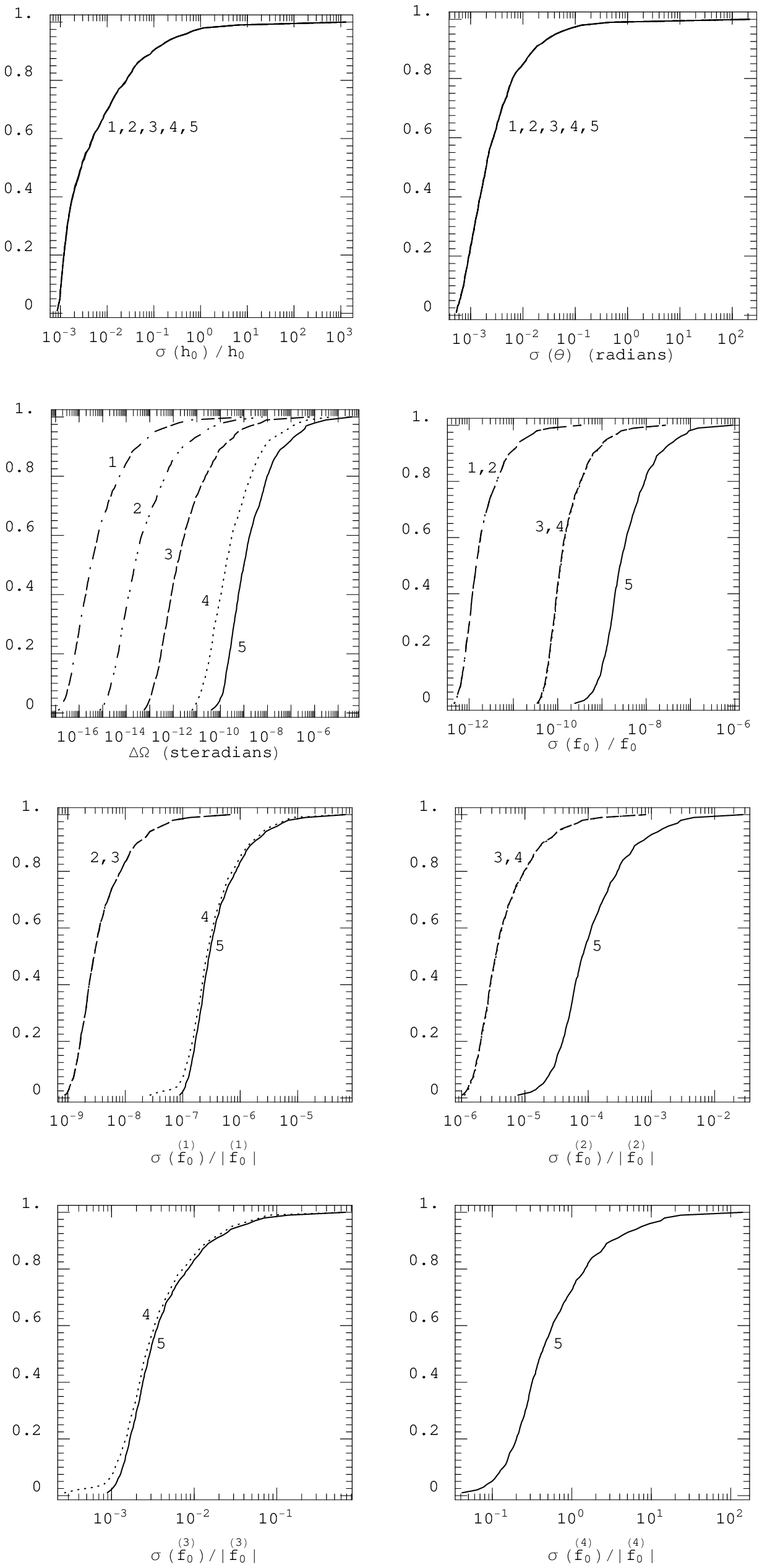}}
\end{picture}
\caption{\label{f:hie}
Cumulative distribution functions of the simulated rms errors of the
signal parameters for all-sky searches.  The observation time $T_o=120$ days.
The advanced Hanford detector is assumed in the simulations.  The neutron star
parameters are the same as in Figure 1.  For these set of parameters the model
of the signal's phase  consistent with the 1/4 of a cycle criterion is obtained 
from Eq.\ (\ref{ap200}) by setting $s_1=4$, $s_2=3$, and $s_3=0$.  This is the 
model number 5 in the plots given here.  Models 1 through 4 are defined as 
follows:  model 4---$s_1=3$, $s_2=3$, $s_3=0$; model 3---$s_1=2$, $s_2=2$, 
$s_3=0$; model 2---$s_1=1$, $s_2=1$, $s_3=0$; model 1---$s_1=0$, $s_2=0$, 
$s_3=0$. We observe a curve merging of the various models that
corresponds exactly to the intersection points of the curves
in Figure 14 for $x_o = 0.5$. This shows that qualitative properties 
of the Fisher matrix for the polynomial model are preserved in the exact model.  
The five cumulative distributions of the amplitude parameters in the 
two top panels are indistinguishable. This confirms an effective 
decorrelation of the amplitude and the phase parameters.}  
\end{figure}\end{center}


\begin{thebibliography}{99}
\bibitem{quartiles}
Quantile $q_x$ gives a value $z_x$ of random variable $z$ such that probability
that $z<z_x$ is less than or equal to $x$.  The quantile values at $x$ = 0.25,
0.5, and 0.75 are called the quartiles.
\end{thebibliography}
\end{document}